\documentclass[12pt]{article}
\usepackage{geometry}
\usepackage{setspace}
\usepackage{fancyhdr}
\usepackage{amsfonts}
\usepackage{amsmath}
\usepackage{verbatim}
\usepackage{listings}
\usepackage{lscape}
\usepackage{graphics}
\usepackage{tabularx}
\usepackage{float}
\usepackage{breqn}
\usepackage{wasysym}
\usepackage{fix-cm}
\usepackage{bbm}
\usepackage{color}
\usepackage{amsthm}
\usepackage{enumitem}
\usepackage{authblk}
\usepackage{mysty}

\usepackage{appendix}

\usepackage{graphicx}

\usepackage{multirow}
\usepackage{parskip}
\usepackage{amsmath, amsthm, amssymb}
\usepackage{dsfont}
\usepackage{bm}
\usepackage{parskip}

\usepackage{framed,enumitem}
\usepackage{tikz}
\usetikzlibrary{positioning}
\usepackage{accents}
\usepackage{comment}
\usepackage[ruled,vlined]{algorithm2e}
\usepackage{breqn}

\usepackage[round]{natbib}

\makeatletter
\newtheorem*{rep@theorem}{\rep@title}
\newcommand{\newreptheorem}[2]{%
\newenvironment{rep#1}[1]{%
 \def\rep@title{#2 \ref{##1}}%
 \begin{rep@theorem}}%
 {\end{rep@theorem}}}
\makeatother

\newreptheorem{mycor}{Corollary}
\newreptheorem{myprop}{Proposition}

\title{Existence and optimisation of the partial correlation graphical lasso}
\author[1,2,3]{Jack Storror Carter}
\author[1]{Cesare Molinari}
\affil[1]{Dept. of Mathematics, University of Genoa, Italy}
\affil[2]{Dept. of Economics and Business, Universitat Pompeu Fabra, Spain}
\affil[3]{Data Science Center, Barcelona Graduate School of Economics, Spain}
\date{}

\begin{document}

\maketitle

\begin{abstract}
The partial correlation graphical LASSO (PCGLASSO) is a penalised likelihood method for Gaussian graphical models which provides scale invariant sparse estimation of the precision matrix and improves upon the popular graphical LASSO method. However, the PCGLASSO suffers from computational challenges due to the non-convexity of its associated optimisation problem. This paper provides some important breakthroughs in the computation of the PCGLASSO.  First, the existence of the PCGLASSO estimate is proven when the sample size is smaller than the dimension - a case in which the maximum likelihood estimate does not exist. This means that the PCGLASSO can be used with any Gaussian data. Second, a new alternating algorithm for computing the PCGLASSO is proposed and implemented in the \texttt{R} package \texttt{PCGLASSO} available at \url{https://github.com/JackStorrorCarter/PCGLASSO}. This was the first publicly available implementation of the PCGLASSO and provides competitive computation time for moderate dimension size.
\end{abstract}

Consider the problem of estimating a sparse $p \times p$ dimensional Gaussian precision matrix $\Theta = (\Theta_{ij})$ based on a sample covariance matrix $S$.  This relates to Gaussian graphical models because zero entries in $\Theta$ correspond to conditional independence relationships between the variables.  However, the maximum likelihood estimate (MLE), $S^{-1}$, is an unstable estimator of $\Theta$ and is not sparse.  A common approach is to instead use a penalised likelihood, the most popular of which is the graphical lasso (GLASSO) \citep{Yuan2007,Banerjee2008,Friedman2008}.  The GLASSO estimates $\Theta$ via an $l_1$ penalised likelihood which is the solution to the optimisation problem
\begin{equation}
\argmax_{\Theta \in \calS} \log( \det( \Theta ) ) - \mathrm{tr}( S \Theta ) 
- \rho \sum_{i=1}^p \sum_{j=1}^p \vert \Theta_{ij} \vert
\end{equation}
where $\rho>0$ is a regularisation parameter and $\calS$ is the space of positive definite matrices.

An important reason for the popularity of the GLASSO is that it requires only the solution to a convex optimisation problem.  This greatly facilitates computation, allowing for calculation of the GLASSO estimate in a matter of seconds, even in high dimensions.  See \cite{Friedman2008} for a method implemented in the \texttt{R} package \texttt{glasso}, and \cite{Sustik2012} for an improved algorithm implemented in the \texttt{R} package \texttt{glassoFast}.  Alternative algorithms include, for example, the P-GLASSO and DP-GLASSO of \cite{Mazumder2012} and the GOLAZO of \cite{lauritzen2022}.

Despite its computational convenience, the GLASSO does have some issues.  One issue in particular, highlighted by \cite{Carter2023}, is that the GLASSO estimate is not invariant to scalar multiplication of the variables.  To remedy this, \cite{Carter2023} proposed an alternative penalised likelihood, called the partial correlation graphical lasso (PCGLASSO), which is based on a reparameterisation of $\Theta$ in terms of the partial correlations.  That is, we write $\Theta = \theta^{1/2}\Delta\theta^{1/2}$ where $\theta$ is the diagonal matrix with diagonal entries $\theta_{ii}=\Theta_{ii}$ and $\Delta = (\Delta_{ij})$ is the matrix with unit diagonal and off-diagonals $\Delta_{ij} = \Theta_{ij}/\sqrt{\Theta_{ii}\Theta_{jj}}$.  The PCGLASSO estimate is the solution to the optimisation problem
\begin{equation*}
\argmax_{\theta_{ii} > 0, \ \Delta \in \calS_1} \log( \det( \Delta ) ) + c \sum_{i} \log( \theta_{ii} ) - \mathrm{tr}\left( S \theta^{\frac{1}{2}} \Delta \theta^{\frac{1}{2}} \right) - \rho \sum_{i \neq j} \vert \Delta_{ij} \vert
\end{equation*}
where $\calS_1$ is the space of positive definite matrices with unit diagonal.  This includes two parameters - $\rho \geq 0$ which controls the $l_1$ penalty on the partial correlations and $c>0$ which controls the logarithmic penalty on the $\theta_{ii}$. Values of $c < 1$ penalise the diagonals, when compared to the log-likelihood function, $c=1$ gives no penalty and $c>1$ works to inflate the diagonal entries. \cite{Carter2023} proposed the use of $c=1-4/n$ where $n$ is the sample size, but here we consider the more general version.  For notational simplicity we use an additional transformation $\xi = \theta^{1/2}$ where the diagonal entries of $\xi$ are equal to $\xi_{i} = \theta_{ii}^{1/2}$ change to a minimisation problem.  The optimisation problem then becomes
\begin{equation}\label{eq:PCGLASSO}
\argmin_{\xi_{ii} > 0, \ \Delta \in \calS_1} - \log( \det( \Delta ) ) - 2 c \sum_{i} \log( \xi_{i} ) + \mathrm{tr}\left( S \xi \Delta \xi \right) + \rho \sum_{i \neq j} \vert \Delta_{ij} \vert
\end{equation}

While the PCGLASSO does not define a convex optimisation problem, it is conditionally convex in $\Delta$ when $\xi$ is fixed (\cite{Carter2023}, Proposition 4). On the other hand, when $\Delta$ is held fixed, the objective function in terms of $\xi$ is differentiable and the problem is convex. This opens the possibility for an alternating algorithm where optimisation of $\Delta$ for fixed $\xi$ benefits from convex optimisation methods and optimisation of $\xi$ for fixed $\Delta$ only requires minimisation of a differentiable function.  In this paper we propose such an algorithm which is implemented in the \texttt{R} package \texttt{PCGLASSO} and is available at \url{https://github.com/JackStorrorCarter/PCGLASSO}.

A further benefit of the GLASSO is that the solution to the optimistation problem exists even when $S$ is not positive definite, but only positive semidefinite. This occurs in Gaussian data when the sample size is less than $p$. While it is trivial to show that the PCGLASSO solution exists when $S$ is positive definite, it was previously unknown whether it exists for positive semidefinite $S$. Here it will be shown that the PCGLASSO solution does exist for any Gaussian data for specific choices of the parameter $c$.

As this paper was being prepared for submission, a new paper was released with important theoretical and computational results for the PCGLASSO \citep{bogdan2025identifying}. Of relevance to this paper are results regarding the uniqueness of the PCGLASSO solution and a new computation method. Specifically, for some $S$ and $\rho$, the optimisation problem (\ref{eq:PCGLASSO}) can have multiple local optima. However, if $S$ is close to a diagonal matrix (i.e. the sample correlations are small), or the penalty parameter $\rho$ is close to 0, then the solution is guaranteed to be unique. Their computation algorithm uses a similar alternating algorithm as proposed in this paper, but with different algorithms for solving the subproblems of optimising $\Delta$ and $\xi$. A comparison to the computation method of this paper was made in \cite{bogdan2025identifying}, Appendix A, apparently showing much faster performance than our propsed method. However, these comparisons do not show the whole picture because they do not consider the objective function values achieved by each method. In fact, as will be shown later in this paper, the apparent speed of the method of \cite{bogdan2025identifying} comes mostly from having a high threshold for convergence, leading the algorithm to terminate while still relatively far from the optimum. In this paper we will perform more thorough comparisons between our proposed method and that of \cite{bogdan2025identifying}, taking into account the objective function value. These comparisons come out in favour of our proposed method, with the currently available implementations of both methods.

The rest of the paper is organised as follows. Section \ref{sec:existence} proves the existence of the solution to the PCGLASSO optimisation problem, even when the sample size is smaller than the problem dimension.  Section \ref{sec:algo} introduces the algorithm and discusses details about initialising the algorithm, parameter choice, stopping rules and convergence.  Section \ref{sec:test} compares its performance to a simple coordinate descent algorithm and to the computation method of \cite{bogdan2025identifying}, and tests its speed in comparison to \texttt{glasso}, \texttt{glassoFast} and another competing penalised likelihood method. Section \ref{sec:disc} concludes with a discussion.

\section{Solution existence}\label{sec:existence}

In a Gaussian sample, when the sample size $n$ is greater than the matrix dimension $p$, the sample covariance matrix $S$ is almost surely positive definite, the MLE exists and is equal to $S^{-1}$, and it is easy to see that the GLASSO and PCGLASSO estimates also exist.  On the other hand, when $n \leq p$, $S$ is not positive definite, but only positive semidefinite with probability $1$.  In this case the MLE does not exist because the objective function tends to infinity as certain eigenvalues of $\Theta$ tend to infinity.  However, an additional benefit of the GLASSO estimate is that it still exists even when $S$ is only positive semi definite, for any choice of the penalty parameter $\rho > 0$ \citetext{\citealp[Theorem 1]{Banerjee2008}; \citealp[Theorem 8.7]{lauritzen2022}; \citealp{carter2025existence}}. This is because the penalty term regularises the objective function, meaning it no longer tends to infinity as eigenvalues of $\Theta$ increase.  The corresponding penalty term in the PCGLASSO on the partial correlations does not have the same effect, because $| \Delta_{ij} | < 1$ and so the penalty term is bounded.  However, for certain choices of the other penalty parameter $c$ on the diagonal entries, the existence of the PCGLASSO estimate is also ensured, even when $n \leq p$.

\begin{theorem}\label{prop}
    Let $X_1,\dots,X_n \overset{\mathrm{iid}}{\sim} N_p(\mu,\Sigma)$ be independent Gaussian random vectors with mean $\mu$ and positive definite covariance matrix $\Sigma$ with $2 \leq n \leq p$. Let $S = \frac{1}{n}(X - \bar{X})(X - \bar{X})^{\T}$ be the sample covariance matrix, where $X = (X_1,\dots,X_n)$ and $\bar{X}$ is the $p \times n$ matrix with columns equal to $\frac{1}{n}(X_1 + \dots + X_n)$.
    
    Then a solution to the PCGLASSO optimisation problem (\ref{eq:PCGLASSO}) with input matrix $S$ exists with probability 1 for any $\rho \geq 0$ and $c < 1 - \frac{k}{p}$ where $k = p-(n-1)$.
\end{theorem}

The proof of this result is given in Appendix \ref{app:proof}.

\begin{figure}[h]
\centering
\begin{tabular}{ccc}
    \includegraphics[scale=0.73]{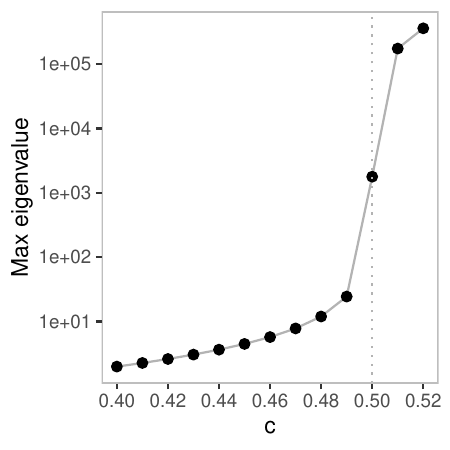} &
    \includegraphics[scale=0.73]{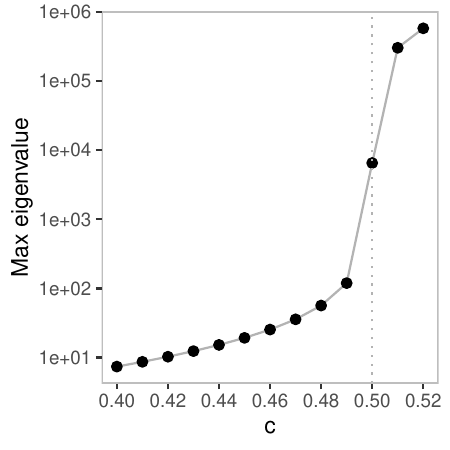} &
    \includegraphics[scale=0.73]{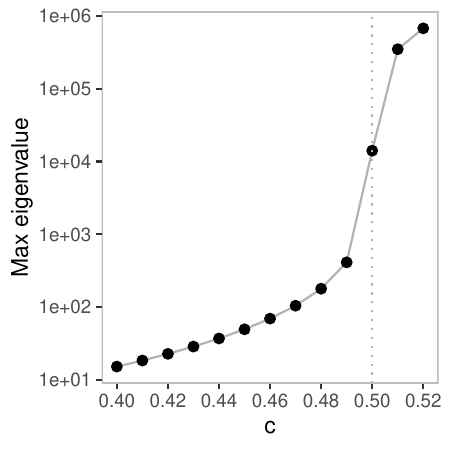}
\end{tabular}
\caption{Maximum eigenvalue of approximate PCGLASSO estimate for $p=2$ setting (left), star graph with $p=10$, $n=6$ (centre) and star graph with $p=20$, $n=11$ (right) for different values of $c$.}
\label{fig:cplot}
\end{figure}

This result proves that the PCGLASSO estimate exists for any $S$ generated from Gaussian data as long as $c < 1 - \frac{k}{p}$. 
Existence is not determined for $c > 1 - \frac{k}{p}$, but using the optimisation algorithm that will be described in Section \ref{sec:algo} with $\rho=0$ we provide empirical evidence that the bound in Theorem \ref{prop} is tight at least in some cases. We consider three settings - $p=2$ with 
$$
S = \begin{pmatrix}
    1 & 1 \\
    1 & 1
\end{pmatrix},
$$
$S$ simulated from a star graph (see Section \ref{sec:test}) with $p=10$, $n=6$ and from a star graph with $p=20$, $n=11$.  In each of these settings Proposition \ref{prop} says that the PCGLASSO solution exists whenever $c < 0.5$. Non-existence of the PCGLASSO solution is characterised by infinite eigenvalues in the solution of $\Theta$. Figure \ref{fig:cplot} shows the maximum eigenvalue of the solution provided by the optimisation algorithm for different values of $c$ (note that even if the true solution has an infinite eigenvalue, the algorithm will terminate early when the objective function is very flat or a maximmum number of iterations is reached). In each case we see a large jump in the maximum eigenvalue at 0.5, suggesting that this is a critical point for the solution existence.

\section{Alternating algorithm}\label{sec:algo}

As discussed, the nature of the PCGLASSO optimisation problem lends itself to an alternating algorithm where $\Delta$ and $\xi$ are optimised in turn while the other is held fixed. This can benefit from the conditional convexity in $\Delta$ and differentiability in $\xi$. As a first step for such an algorithm, we investigate the subproblems of optimising $\Delta$ and $\xi$ while the other is held fixed.

\subsection{Optimisation of $\Delta$}\label{subsec:DR}

First we consider optimisation of $\Delta$ when $\xi$ is fixed.  By writing $\tilde{S} = \xi S \xi$, the optimisation problem is
\begin{equation}\label{eq:deltaop}
\argmin_{\Delta \in \calS_1} -\log( \det( \Delta ) ) + \mathrm{tr}\left( \tilde{S} \Delta \right) + \rho \sum_{i \neq j} \vert \Delta_{ij} \vert
\end{equation}
To numerically solve (\ref{eq:deltaop}) we propose using a Douglas-Rachford splitting (DRS) algorithm \citep{douglas1956numerical}. Consider the two functions
\begin{align*}
    f(\Delta) = -\log \det (\Delta) + \iota_{\calS}(\Delta), &&
    g(\Delta) = \rho \sum_{i\neq j} \abs{\Delta_{ij}} + \mathrm{tr}\left( \tilde{S} \Delta \right) + \iota_{M_1}(\Delta)
\end{align*}
where $M_1$ is the set of matrices with unit diagonal. For a set $A$, the function $\iota_A(x)$ is equal to 0 if $x \in A$ and $\infty$ otherwise. This enforces the constraints on $\Delta$ - $f$ is equal to $\infty$ if $\Delta$ is not positive definite and $g$ is equal to $\infty$ if $\Delta$ has non-unit diagonal.  The optimisation problem (\ref{eq:deltaop}) is equivalent to
\begin{equation*}
    \min_{(\Delta,\tilde{\Delta}) \in V} \quad f(\Delta) + g(\tilde{\Delta}),
\end{equation*}
where $V=\{(\Delta,\tilde{\Delta})\,:\, \Delta=\tilde{\Delta}\}$.

Given starting values $x^{(0)},y^{(0)},z^{(0)}$, the DRS algorithm for $k=0,1,2,...$ iterates over the following updates,
\begin{align*}
    x^{(k+1)} &= x^{(k)} + \lambda(z^{(k)} - y^{(k)}) \\
    y^{(k+1)} &= \prox_{\alpha f}(x^{(k+1)}) \\
    z^{(k+1)} &= \prox_{\alpha g}(2y^{(k+1)} - x^{(k+1)})
\end{align*}
where $\prox$ is the proximal point operator, $\alpha$ is the proximal step size and $\lambda$ is the relaxation parameter.  Generally $\lambda$ is allowed to depend on $k$, but for simplicity we only consider fixed $\lambda$.  For a proper selection of the parameters $\alpha$ and $\lambda$, both $y^{(k)}$ and $z^{(k)}$ converge to a solution of the optimisation problem (\ref{eq:deltaop}).

The proximal point operator for $g$ is
\begin{equation*}
        \left(\prox_{\alpha g}(\Delta)\right)_{ij} = \begin{cases} 1& i = j\\ \mathrm{shrink}(\Delta_{ij}-\alpha \tilde{S}_{ij},\, \alpha \rho) & i \neq j \end{cases}
\end{equation*}
where $\mathrm{shrink}(a,b)$ shrinks $a$ towards $0$ by the amount $b$. See Appendix \ref{app:prox} for details on the derivation of the proximal point operators.

For $\prox_{\alpha f}$ we consider the spectral decomposition $\Delta = V \Sigma V^{\T}$ where $\Sigma=\diag(\sigma_1,...,\sigma_p)$ is the diagonal matrix of eigenvalues of $\Delta$ and $V$ is the matrix with columns equal to the eigenvectors of $\Delta$.  Then the proximal point operator is (see Appendix \ref{app:prox} for details)
\begin{equation*}
    \prox_{\alpha f}\left( \Delta \right)=V \tilde{\Sigma} V^{\T}.
\end{equation*}
where $\tilde{\Sigma}$ is the diagonal matrix with entries
\begin{equation*}
\begin{split}
        \tilde{\sigma}_{i}
        & = \frac{1}{2}\left(\sigma_i+\sqrt{\sigma_i^2+4\alpha}\right) >0.
\end{split}
\end{equation*}

Although both sequences $y^{(k)}$ and $z^{(k)}$ converge to the solution of (\ref{eq:deltaop}), we choose to output the value of $z^{(k)}$ after sufficient convergence. This is because for finite $k$, $y^{(k)}$ is guaranteed to be positive definite while $z^{(k)}$ is guaranteed to have unit diagonal.  After sufficient convergence, $z^{(k)}$ is also guaranteed to be positive definite, but $y^{(k)}$ is not guaranteed to have unit diagonal for finite $k$.  Hence outputting $z^{(k)}$ ensures that both conditions are satisfied.

The steps of the DRS algorithm are summarised in Algorithm \ref{alg:drdelta}.

\begin{algorithm}[H]\label{alg:drdelta}
\SetAlgoLined
Select starting points $x^{(0)},y^{(0)},z^{(0)}$ and parameters $\alpha,\lambda$.  For $k=0,1,2,...$ update $x^{(k)},y^{(k)},z^{(k)}$ as follows:
\begin{enumerate}
    \item Set $x^{(k+1)} = x^{(k)} + \lambda (z^{(k)} - y^{(k)})$.
    \item Let $V \Sigma V^{\T}$ be the spectral decomposition of $x^{(k+1)}$ where $\Sigma$ has diagonal entries $\sigma_1,...,\sigma_p$.
    \item Let $\tilde{\Sigma}$ be the diagonal matrix with diagonal entries $\tilde{\sigma}_i = \frac{1}{2}\left(\sigma_i+\sqrt{\sigma_i^2+4\alpha}\right)$, $i=1,...,p$ and set $y^{(k+1)} = V \tilde{\Sigma} V^{\T}$.
    \item Set $z^{(k+1)}$ with diagonal entries $z^{(k+1)}_{ii} = 1$, $i=1,...,p$ and off-diagonals $z^{(k+1)}_{ij} = \mathrm{shrink}(2 y^{(k+1)}_{ij} - x^{(k+1)}_{ij} - \alpha \tilde{S}_{ij}, \alpha \rho)$ for $i \neq j$.
    \item When some stopping rule is achieved, stop and return $\Delta = z^{(k+1)}$ (and $x = x^{(k+1)}, y = y^{(k+1)}$ if required).
\end{enumerate}
 \caption{Optimisation of $\Delta$ - DRS algorithm.}
\end{algorithm}

\subsection{Optimisation of $\xi$}\label{subsec:FB}

Next we consider optimisation of $\xi$ when $\Delta$ is held fixed. Removing constants, the optimisation problem can be written as
\begin{equation}\label{eq:ximin}
\argmin_{\xi > 0}  \mathrm{tr}\left( S \xi \Delta \xi \right) - 2 c \sum_{i} \log( \xi_i )
\end{equation}
While this objective function is relatively simple, the interaction term between different $\xi_i$ in the trace term, $\mathrm{tr}\left( S \xi \Delta \xi \right) = \sum_{i,j} S_{ij} \Delta_{ij} \xi_i \xi_j$, inhibits an analytic solution. By taking the derivative of the objective function with respect to $\xi_i$, we find that the optimality conditions are
\begin{equation*}
    2 \sum_{j} S_{ij} \Delta_{ij} \xi_j - \frac{c}{\xi_i}=0.
\end{equation*}
Instead we propose using forward-backward splitting (FBS) \citep{combettes2011proximal} to numerically find the optimal solution.

Starting from $\xi^{(0)}$, for $k=0,1,2,...$ the FBS algorithm updates $\xi$ as
\begin{equation*}
    \xi^{(k+1)}=\prox_{\gamma \left(-2c\sum_{i=1}^p \log(\xi_{i}) + \iota_{\R^p_+}(\xi_1,\dots,\xi_p)\right)} \left(\xi^{(k)} - \gamma \nabla \left[ \mathrm{tr}\left( S \xi \Delta \xi \right) \right](\xi^{(k)})\right),
\end{equation*}
where
$$ \nabla_i \left[ \mathrm{tr}\left( S \xi \Delta \xi \right) \right](\xi) = 2 \sum_{j} S_{ij} \Delta_{ij} \xi_j$$
and (see Appendix \ref{app:prox} for details)
\begin{equation*}
    \left[\prox_{\gamma \left(-2c\sum_{i=1}^p \log(\xi_{i}) + \iota_{\R^p_+}(\xi_1,\dots,\xi_p)\right)}(\xi)\right]_i = \frac{1}{2}\left[\xi_i+\sqrt{\xi_i^2+8c\gamma}\right].
\end{equation*}

This is summarised more concisely in Algorithm \ref{alg:fbxi}.

\begin{algorithm}[H]\label{alg:fbxi}
\SetAlgoLined
Select starting point $\xi^{(0)}$ and parameter $\gamma$.  For $k=0,1,2,...$ update $\xi^{(k)}$ as follows:
\begin{enumerate}
    \item For $i=1,...,p$, set $\tilde{\xi}^{(k)}_i = \xi^{(k)}_i - 2 \gamma \sum_{j} S_{ij} \Delta_{ij} \xi^{(k)}_j$.
    \item For $i=1,...,p$, set $\xi^{(k+1)}_i = \frac{1}{2}\left[\tilde{\xi}^{(k)}_i+\sqrt{\left(\tilde{\xi}^{(k)}_i\right)^2+8c\gamma}\right]$
    \item When some stopping rule is achieved, stop and return $\xi = \xi^{(k+1)}$.
\end{enumerate}
 \caption{Optimisation of $\xi$ - FBS algorithm.}
\end{algorithm}

\subsection{Final algorithm}

The DRS algorithm for optimising $\Delta$ and the FBS algorithm for optimising $\xi$ can be combined into a single alternating algorithm (see Algorithm \ref{alg:alternating}). A main aspect to note is that the output values $x^{(k)}, y^{(k)}, z^{(k)}$ of the previous DRS run are used as the start points of the next DRS run, rather than resetting then at each run (another option, for example, would be $x^{(k+1)} = y^{(k+1)} = z^{(k+1)} = z^{(k)}$). While not necessary for convergence, we found that this made a great difference to the speed of the algorithm.

\begin{algorithm}[H]\label{alg:alternating}
\SetAlgoLined
Choose start points $x^{(0)} = y^{(0)} = z^{(0)}$ and $\xi^{(0)}$.  For $k=0,1,2,...$, update $x^{(k)}, y^{(k)}, z^{(k)}, \xi^{(k)}$ as follows:
\begin{enumerate}
    \item Choose parameters $\alpha,\lambda$
    \item Set $x^{(k+1)}, y^{(k+1)}, z^{(k+1)}$ using Algorithm \ref{alg:drdelta} with starting points $x^{(k)}, y^{(k)}, z^{(k)}$, parameters $\alpha,\lambda$ and $\tilde{S} = \xi^{(k)} S \xi^{(k)}$.
    \item Choose parameter $\gamma$.
    \item Set $\xi^{(k+1)}$ using Algorithm \ref{alg:fbxi} with starting point $\xi^{(k)}$, parameter $\gamma$ and $\Delta = z^{(k+1)}$.
    \item When some stopping rule is achieved, stop and return $\Theta = \xi^{(k+1)} z^{(k+1)} \xi^{(k+1)}$
\end{enumerate}
 \caption{Alternating algorithm}
\end{algorithm}

%This algorithm requires stopping rules for all three algorithms.  One choice of stopping rules for Algorithms \ref{alg:drdelta} and \ref{alg:fbxi} is to just run them for a single step, i.e. stop after $k=0$.

%\begin{algorithm}[H]\label{alg:onestep}
%\SetAlgoLined
%Choose start points $\Delta^{(0)},\tilde{\Delta}^{(0)},\xi^{(0)}$ and parameters $\gamma_{\textrm{DR}},\gamma_{\textrm{FB}}$.  For $k=0,1,2,...$, update $\Delta^{(k)},\tilde{\Delta}^{(k)},\xi^{(k)}$ as follows:
%\begin{enumerate}
%    \item Set $\Delta^{(k+1)},\tilde{\Delta}^{(k+1)}$ using one iteration of Algorithm \ref{alg:drdelta} with starting points $\Delta^{(k)},\tilde{\Delta}^{(k)}$, parameter $\gamma_{\textrm{DR}}$ and $\tilde{S} = \xi^{(k)} S \xi^{(k)}$.
%    \item Set $\xi^{(k+1)}$ using one iteration of Algorithm \ref{alg:fbxi} with starting point $\xi^{(k)}$, parameter $\gamma_{\textrm{FB}}$ and $\Delta = \Delta^{(k+1)}$.
%    \item When some stopping rule is achieved, stop and return $\Theta = \xi^{(k+1)} \Delta^{(k+1)} \xi^{(k+1)}$
%\end{enumerate}
% \caption{One step alternating algorithm}
%\end{algorithm}

\subsection{Starting values, parameter choice and stopping rules}\label{subsec:parameters}

Algorithm \ref{alg:alternating} requires the choice of starting points $x^{(0)}, \xi^{(0)}$, parameter values $\alpha,\lambda,\gamma$ and stopping rules for each of Algorithms \ref{alg:drdelta}, \ref{alg:fbxi} and \ref{alg:alternating}.  Both the speed and convergence of the algorithm depend on these choices.

For starting points, an obvious choice is using the inverse of $S$ when it is positive definite.  That is, we choose $x^{(0)}, \xi^{(0)}$ such that $\xi^{(0)} x^{(0)} \xi^{(0)} = S^{-1}$.  When $S$ is not positive definite, using the eigendecomposition of $S$ we add a small amount to each of the eigenvalues to ensure they are all positive.  The starting point is then obtained from the inverse of this new positive definite matrix.  We add an amount to the eigenvalues such that the minimum eigenvalue is equal to 1.  When there is a sequence of penalty parameters $\rho_1 < ... < \rho_k$, the optimised $\Delta$ and $\xi$ for the previous penalty parameter $\rho_{i-1}$ can be used as the starting point of the next penalty parameter $\rho_i$.

While the DRS algorithm is guaranteed to converge for any $\lambda \in (0,2)$ \citep{combettes2004solving}, the choice of parameters $\alpha,\lambda$ can have a large effect on the speed of convergence with the optimal values depending on the levels of strong convexity and smoothness of $f(\Delta) = -\log \det (\Delta) + \iota_{\calS}(\Delta)$.  Since $f(\Delta)$ is neither strongly convex nor smooth, it was shown by \citet{seidman2019control} that an optimal choice is $\lambda=1$. We use this as a default choice along with $\alpha=1$.  These values remain fixed across each iteration of Algorithm \ref{alg:alternating}.

When the FBS parameter $ \gamma < 2/L $, where $L$ is a Lipschitz constant of $\nabla h(\xi)$ with $h(\xi) = \mathrm{tr}\left( S \xi \Delta \xi \right)$, the FBS is guaranteed to converge \citep{goldstein2014field}. $L$ is a Lipschitz constant of $\nabla h(\xi)$ if and only if
$$ \lVert \nabla h(\xi) - \nabla h(\xi') \rVert_2 \leq L \lVert \xi - \xi' \rVert_2 $$
for all $\xi,\xi'$. Since $ \nabla h(\xi) = 2 (\Delta \cdot S) \xi $,
\begin{align*}
   \lVert \nabla h(\xi) - \nabla h(\xi') \rVert_2 &= \lVert 2 (\Delta \cdot S) \xi - 2 (\Delta \cdot S) \xi' \rVert_2 \\
   & \leq 2 \lVert \Delta \cdot S \rVert_2 \lVert \xi - \xi' \rVert_2 \\
   &= 2 \sigma_{\textrm{max}}(\Delta \cdot S ) \lVert \xi - \xi' \rVert_2
\end{align*}
where $\sigma_{\textrm{max}}(\Delta \cdot S )$ denotes the largest eigenvalue of $\Delta \cdot S$. Hence $L = 2 \sigma_{\textrm{max}}(\Delta \cdot S )$ is a Lipschitz constant and taking $ \gamma < 1/\sigma_{\textrm{max}}(\Delta \cdot S )$ guarantees convergence. We choose a value of $\gamma = 0.9/\sigma_{\textrm{max}}(\Delta \cdot S )$. This value depends on the current value for $\Delta$ and so must be updated at each iteration of Algorithm \ref{alg:alternating}.

For the stopping rule of Algorithm \ref{alg:alternating}, we consider the sum of the absolute parameter changes of the two parameters $\Delta,\xi$. These sums are normalised by the sum of absolute values of the previous iteration parameters to make the stopping rule more consistent over dimensions $p$ and scales, and to ensure that the algorithm does not terminate early when there is sparsity in $\Delta$. The stopping rule for threshold $t$ is then 
\begin{equation}\label{eq:StoppingRule}
    \frac{ \sum_{i \neq j} \abs{ \Delta_{ij}^{(k+1)} - \Delta_{ij}^{(k)} } }{\max\{\sum_{i \neq j} \abs{ \Delta_{ij}^{(k)} }, 10^{-8} \} } + \frac{ \sum_{i} \abs{ \xi_{i}^{(k+1)} - \xi_{i}^{(k)} } }{\sum_{i} \abs{ \xi_{i}^{(k)} } } < t
\end{equation}
The denominator of the $\Delta$ condition is also given a lower bound to account for when $\Delta^{(k)}$ is a diagonal matrix.

Similar stopping rules are used for the DRS and FBS algorithms with the $\Delta$ part of (\ref{eq:StoppingRule}) used for the DRS and the $\xi$ part used for the FBS. We denote the thresholds for the two algorithms by $t_{DR}$ and $t_{FB}$. Before terminating the DRS algorithm it is confirmed that the current $\Delta$ value is positive definite. Additional checks are also made that the objective function value has not increased, since both the DRS and FBS algorithms can move to a worse point before moving towards the optimum.

In our experience, the overall level of convergence is largely controlled by the choice of $t$. This choice will be explored further in Section \ref{subsec:PCfast}, but as a default we use $t=10^{-4}$ which provides a balance between speed and convergence. On the other hand, the choices of $t_{DR},t_{FB}$ have interesting implications on the speed of the algorithm. The theoretical convergence of the overall algorithm depends on the full convergence of the DRS and FBS (see Section \ref{subsec:convergence}), which would suggest choosing small values for $t_{DR},t_{FB}$. Small values mean that each subproblem reaches the near optimal value. This is useful towards the end of the alternating algorithm, when already close to the global optimum. However, at the beginning of the algorithm, when still far from the global optimum, it results in over optimisation of the subproblems leading to a great increase in computation time with little benefit to convergence. Instead, here it is better to have a larger threshold for many fast alternating steps for quicker convergence towards the global optimum. To balance this, we decrease $t_{DR},t_{FB}$ at each iteration of Algorithm \ref{alg:alternating}. To motivate how we decrease them, we have found that $t_{DR},t_{FB} < t$ is generally required for the algorithm to terminate in a reasonable amount of time. Meanwhile, a choice of $t=10^{-3}$ generally ensures the algorithm arrives close to the optimum value in a fast time (see Section \ref{subsec:PCfast}). Hence we initialise $t_{DR},t_{FB}$ at $10^{-3}$ and decrease them at each step until they reach $10^{-1} \times t$. This is done via $$t_{DR} = t_{FB} = \max\{ 10^{-3} \times 0.9^k, 10^{-1} \times t \}$$
where $k$ is the current iteration of Algorithm \ref{alg:alternating}.
%For the stopping rule of Algorithm \ref{alg:alternating}, we use the change in 2 norm at each iteration. 
%as well as the change in the objective function value from Equation (\ref{eq:PCGLASSO}). This is to avoid early termination of the algorithm if an update results in a small change in parameter values but large change in objective function, or vice versa. Specifically, the stopping rule is

%$$ \frac{ \lVert \Delta_{k+1} - \Delta_{k} \rVert_2 }{ \lVert \Delta_{k} \rVert_2 } + \frac{ \lVert \xi_{k+1} - \xi_{k} \rVert_2 }{ \lVert \xi_{k} \rVert_2 } 
%+ \lvert F(\Delta_{k+1},\xi_{k+1}) - F(\Delta_{k},\xi_{k}) \rvert 
%< t$$

%where %$F(\Delta,\xi)$ is the objective function and 
%$t$ is some threshold.  The changes are normalised by the 2 norm of the previous value to make the stopping rule applicable for any dimension $p$, and for ensuring that the algorithm does not terminate early when there is sparsity in the current $\Delta$ value.
%Additionally, because it is possible for the algorithm to move to smaller objective function value between iterations, we also require that $F(\Delta_{k+1},\xi_{k+1}) \geq F(\Delta_{k},\xi_{k})$.
%For the DRS and FBS algorithms we consider the change in 2 norm for $\Delta$ and $\xi$ respectively.  For thresholds we use $t=10^{-5}$ for Algorithm \ref{alg:alternating} and $t=10^{-6}$ for the DRS and FBS algorithms.

\subsection{Convergence}\label{subsec:convergence}

Alternating algorithms for the PCGLASSO problem are guaranteed to converge to a stationary point when the optimisation of the sub-problems is exact \citep[Corollary 14.8]{beck2017first}. This applies, for example, to the coordinate descent algorithm used in \cite{Carter2023}. If the DRS and FBS algorithms return the exact optimisers for their subproblems, then this result also applies to Algorithm \ref{alg:alternating}. In practice this is not guaranteed because the use of stopping rules might mean that the DRS and FBS algorithms terminate before convergence is reached, however it still provides good evidence that Algorithm \ref{alg:alternating} will converge to a stationary point. This is further backed up by the empirical tests in Section \ref{subsec:CD} where Algorithm \ref{alg:alternating} achieves a smaller objective function value and seems to be converging to the same objective function value as the coordinate descent algorithm.

\section{Testing}\label{sec:test}

We test the speed of the algorithm in the star graph setting where the data generating $\Theta$ has diagonals $\theta_{ii}=1$, off-diagonals in the first row and column equal to $\theta_{1i} = \theta_{i1} = -1 / \sqrt{p}$ and all other $\theta_{ij} = 0$.  For a given dimension $p$ and sample size $n$ the data is generated from a multivariate Normal distribution with mean $0$ and covariance $\Theta^{-1}$ to give a sample covariance matrix $S$, which is then standardised to have unit diagonal (i.e. the sample correlation matrix).  For Sections \ref{subsec:CD} and \ref{subsec:glasso}, tests were also conducted for a hub graph, AR2 model and random graph - these results are shown in Appendix \ref{app:furthertesting} and closely match those of the star graph, but with much faster computation times for all methods suggesting that the star graph is a particularly challenging setting.

All testing was done in R version 4.3.1 on a 2022 MacBook Pro with Apple M2 chip running macOS 15.7.1.

\subsection{Comparison to coordinate descent}\label{subsec:CD}

\begin{figure}[]
\centering
\includegraphics[scale=0.7]{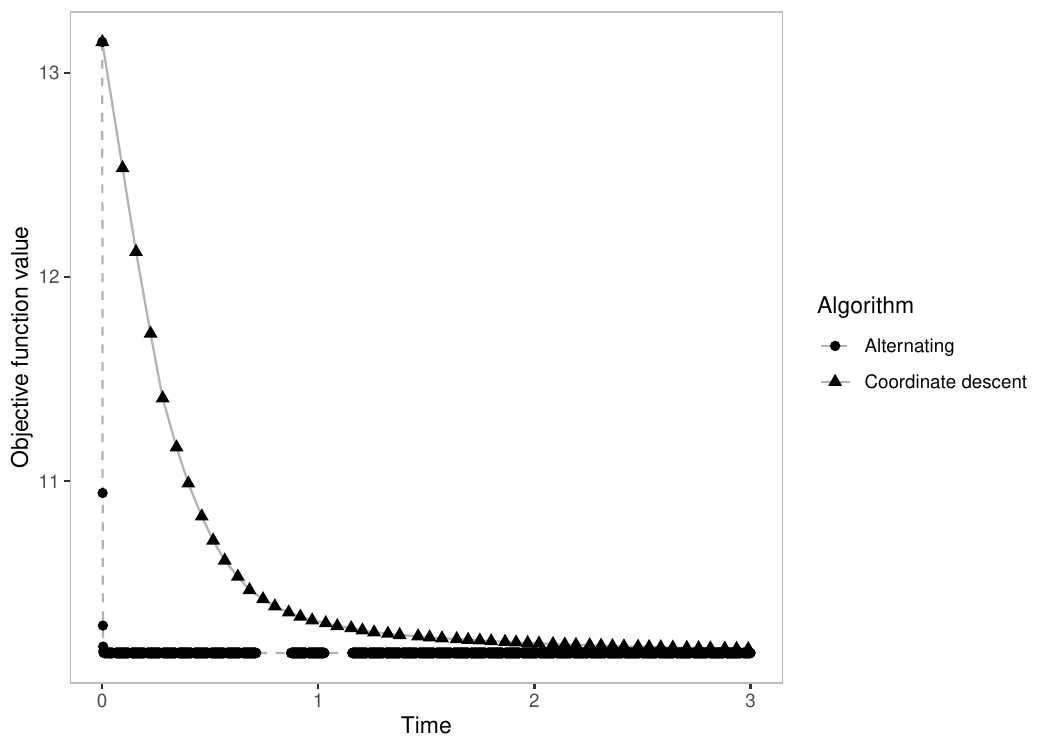}
\caption{Comparison of the proposed alternating algorithm to a coordinate descent algorithm for the star graph with $p=20$, $n=40$. Points correspond to each iteration of the algorithms.}
\label{fig:StarCD}
\end{figure}

We first compare the proposed algorithm to the coordinate descent algorithm used in \cite{Carter2023}.  Figure \ref{fig:StarCD} shows this comparison for a single simulation from the star graph with $p=20$ and $n=40$.  Both algorithms seem to converge towards the same objective function value, however the proposed alternating algorithm has much quicker convergence.  Repeating this experiment 10 times, the alternating algorithm always reached a smaller objective function value than the coordinate descent and always in a much shorter time.

\subsection{Comparison to GLASSO}\label{subsec:glasso}

\begin{figure}[hp]
\centering
\begin{tabular}{c}
	\includegraphics[scale=0.6]{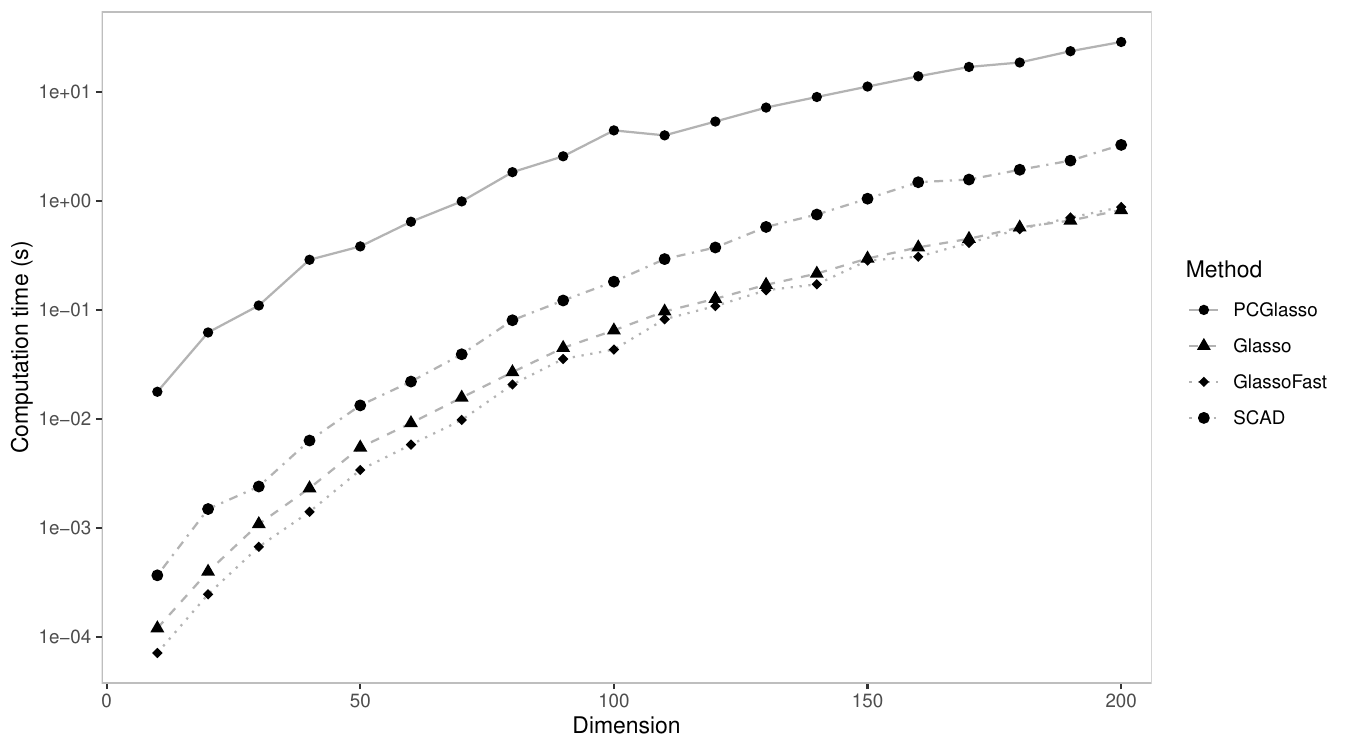} \\
	\includegraphics[scale=0.6]{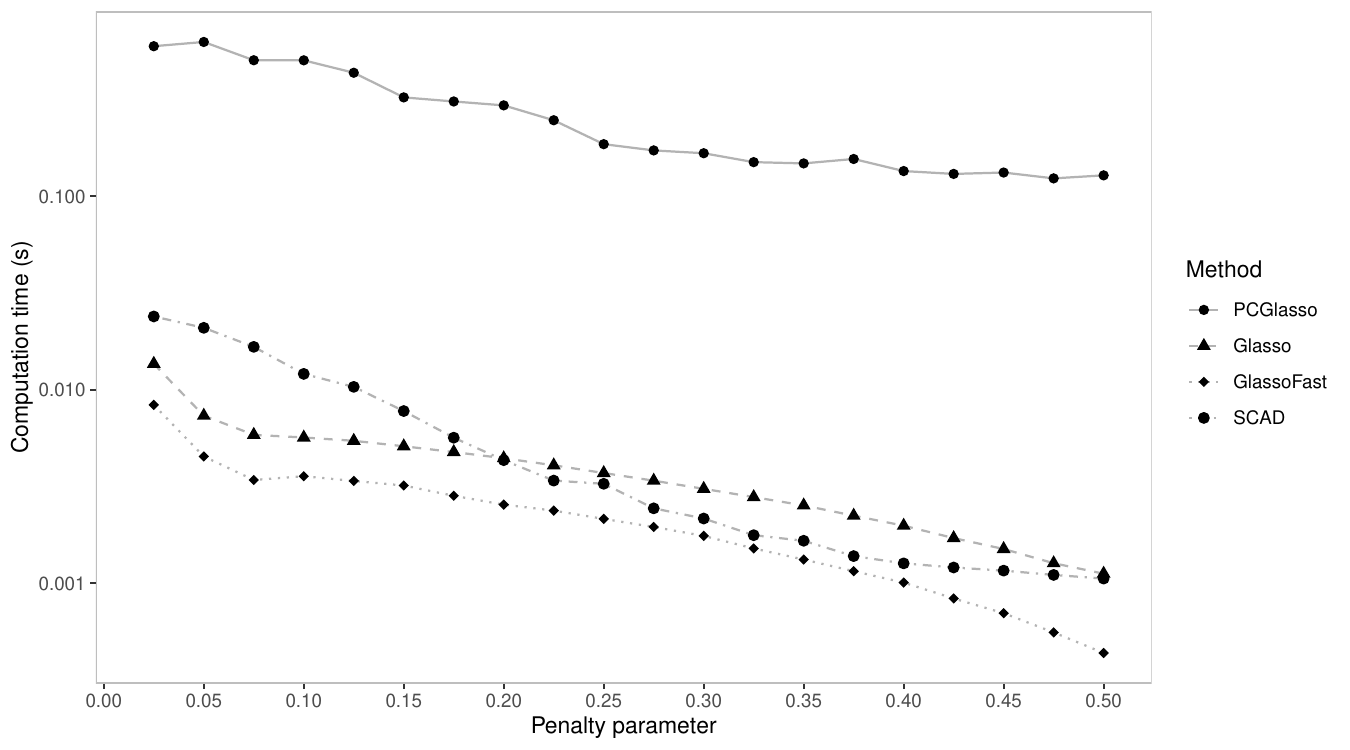} \\
	\includegraphics[scale=0.6]{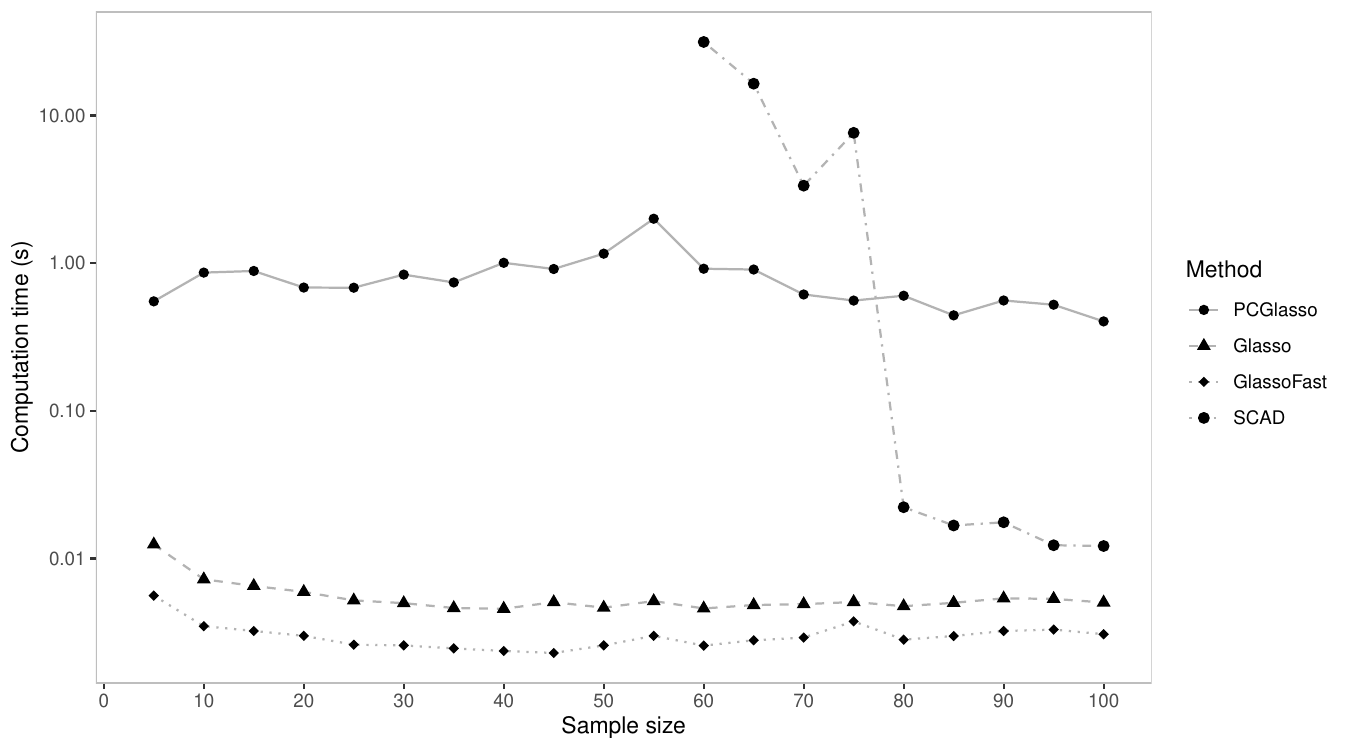}
\end{tabular}
\caption{Comparison of mean computation time for varying dimension (top), penalty parameter (middle) and sample size (bottom) for the star graph.}
\label{fig:Star}
\end{figure}

We now compare the speed of the proposed alternating algorithm for the PCGLASSO to algorithms for other penalised likelihood methods - namely the GLASSO and SCAD \citep{fan2009network} penalties.  The GLASSO is implemented using the R packages \texttt{glasso}, which impletements the method of \cite{Friedman2008}, and \texttt{glassoFast} of \cite{Sustik2012}.  The SCAD penalty is implemented using the package \texttt{GGMncv} \citep{williams2020beyond}.  The GLASSO optimisation problem is a much simpler problem than the PCGLASSO - in fact it is analagous to a single optimisation of $\Delta$, as in Algorithm \ref{alg:drdelta}, without the unit diagonal constraint.  Since we use an alternating algorithm for the PCGLASSO problem, it should be expected to be slower than the GLASSO. The average computation time of each method over 10 replications is shown in Figure \ref{fig:Star} for varying dimension, penalty parameter and sample size.

Results for varying dimension $p = 10,20,\dots,200$ with sample size $n=2p$ and all methods using a penalty parameter of $\rho = 0.1$ (and PCGLASSO also with $c=1$) are in the top panel of Figure \ref{fig:Star}.  We see that, while the computation time of PCGLASSO is slower than the other methods, it scales with dimension in an almost identical, sub-exponential manner.

Results for different values for the penalty parameter $\rho=0.025,0.05,\dots,0.5$ (with $c=1$ fixed for PCGLASSO), with fixed dimension $p=50$ and sample size $n=100$ are in the middle panel of Figure \ref{fig:Star}. All methods have faster computation for larger penalty parameter values. This might be because the objective function becomes more peaked for larger penalty, aiding convergence.

Finally results for fixed dimension $p=50$ and penalty parameter $\rho=0.1$, but with varying sample size $n=5,10,\dots,100$ are in the bottom panel of Figure \ref{fig:Star}. Recall that when $n \leq p$, the sample covariance matrix $S$ is only positive semi definite. In this case the SCAD estimate does not exist, and so for SCAD we only consider $n \geq 60$.  For PCGLASSO, the additional parameter $c$ is chosen based on the threshold given in Proposition \ref{prop} as $c = 0.9(1 - k/p)$.  The speed of PCGLASSO and GLASSO is not impacted too much by the sample size. However, SCAD sees a large change, speeding up as the sample size increases.

\subsection{Comparison to pcglassoFast}\label{subsec:PCfast}

Finally we compare our method to that of \cite{bogdan2025identifying}, which is available in the R package \texttt{pcglassoFast} at \url{https://github.com/PrzeChoj/pcglassoFast}. For clarity we refer to the two methods by their package names - pcglasso for our method and pcglassoFast for that of \cite{bogdan2025identifying}. For both methods the default settings will be used, except for the threshold (referred to as tolerance by pcglassoFast). We will investigate the speed and convergence of both methods for varying thresholds.

\begin{figure}[]
\centering
\includegraphics[scale=0.7]{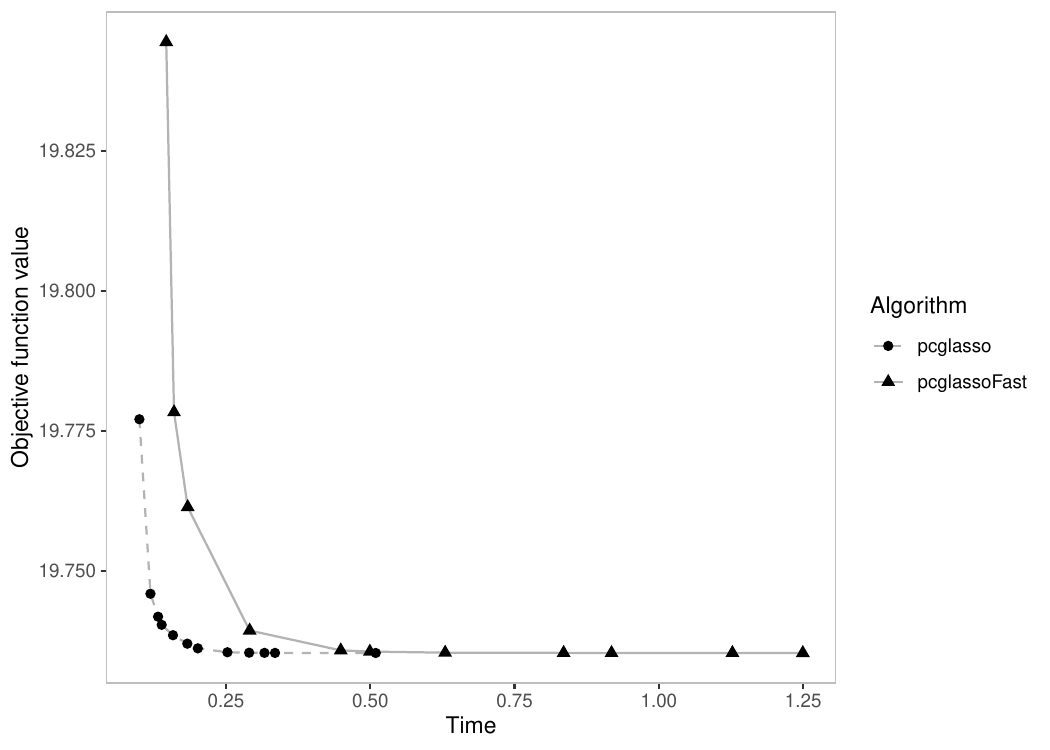}
\caption{Comparison of time and objective function values between pcglasso and pcglassoFast for a star graph $p=50$, $n=100$ example. Points correspond to different threshold/tolerance values that cause the algorithms to terminate at different levels of convergence.}
\label{fig:fasttest1}
\end{figure}

A comparison of pcglasso and pcglassFast has already been conducted in \cite{bogdan2025identifying}, Appendix 1.\footnote{This comparison used an older version of our pcglasso implementation. This has since been updated with minor efficiency improvements and new stopping rules. However, the conclusions of this section are valid for both versions.} This reported much faster computation times for pcglassoFast. However, this was not a fair comparison because they did not consider objective function values. For an example of why this can lead to misleading results, we simulate a single sample covariance matrix from the star graph setting with $p=50$, $n=100$ and use penalty parameters $\rho=0.1$, $c=1$.\footnote{The pcglassoFast function instead uses the parameters $\lambda = \rho$ and $\alpha = 1 - c$.} We run pcglasso and pcglassoFast for a range of different threshold values and record the computation time and objective function value. The results of this are displayed in Figure \ref{fig:fasttest1}. We see that both methods seem to converge to the same objective function value with pcglasso arriving to lower values faster than pcglassoFast. The default threshold value for pcglasso is the rightmost of these points, only terminating after suitable convergence has been achieved. The default threshold for pcglassoFast is the leftmost point, terminating in a quicker time than the default for pcglasso, but still far from the optimum. In fact, the rightmost point for pcglassoFast is the first time it achives a lower objective function value than the default pcglasso, taking over double the time to reach it. This demonstrates why comparing the two methods with their default threshold values is not a fair comparison.

For a more thorough comparison of the two methods we repeat the analysis of Section \ref{subsec:glasso} with smaller ranges of dimension $p \in \{ 10, 50, 100, 150, 200 \}$, penalty parameter $\rho \in \{ 0.01, 0.05, 0.1 \}$ and samples size $n \in \{ 5, 50, 100, 200 \}$. We tested a larger range of threshold values, but for demonstration we display results for three settings that are indicative of the performance of the two methods:
\begin{itemize}
    \item Fast setting - tolerance values that prioritise speed over convergence. For pcglasso $t = 10^{-3}$ and for pcglassoFast $t = 10^{-3}$ (the current default for pcglassoFast).
    \item Balanced setting - tolerance values that balance speed with suitable convergence. For pcglasso $t = 10^{-4}$ (our proposed default value) and for pcglassoFast $t = 10^{-5}$.
    \item Convergence setting - tolerance values that prioritise convergence to the optimum value, at the expense of additional computation time. For pcglasso $t = 10^{-6}$ and for pcglassoFast $t = 10^{-8}$.
\end{itemize}
We also use a 'full convergence' setting for pcglasso with $t = 10^{-8}$ to approximate the optimum objective function value. This threshold value will be used to compare the objective function values that the other settings achieve. Note that in almost every simulated example, the full convergence setting did achieve the lowest objective function value - in particular, less than any achieved by pcglassoFast. However, there was one example when this was not always the case, which will be discussed.

\begin{figure}[hp]
\centering
\begin{tabular}{cc}
	\includegraphics[scale=0.55]{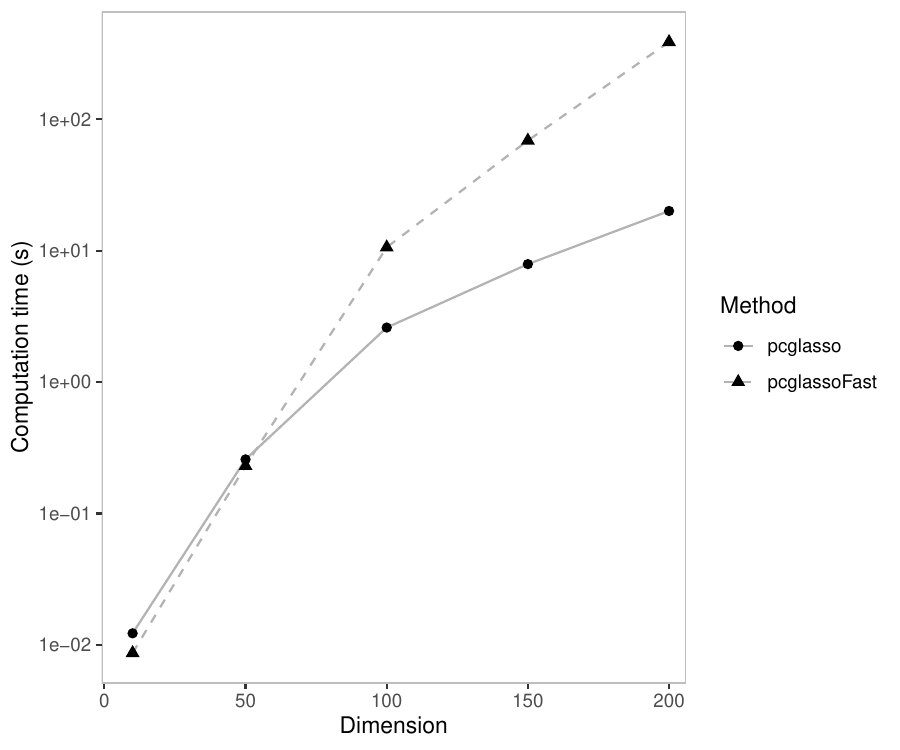} & \includegraphics[scale=0.55]{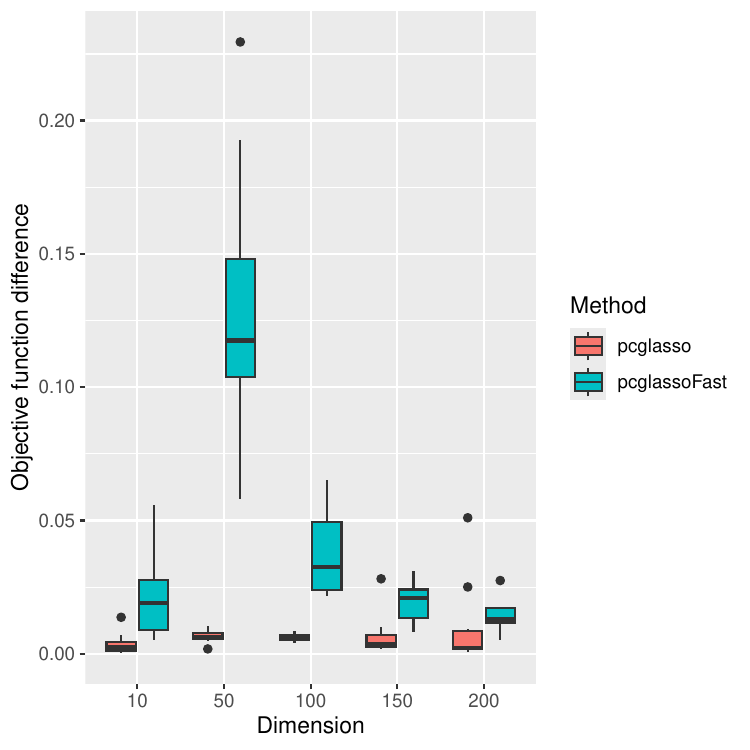} \\
	\includegraphics[scale=0.55]{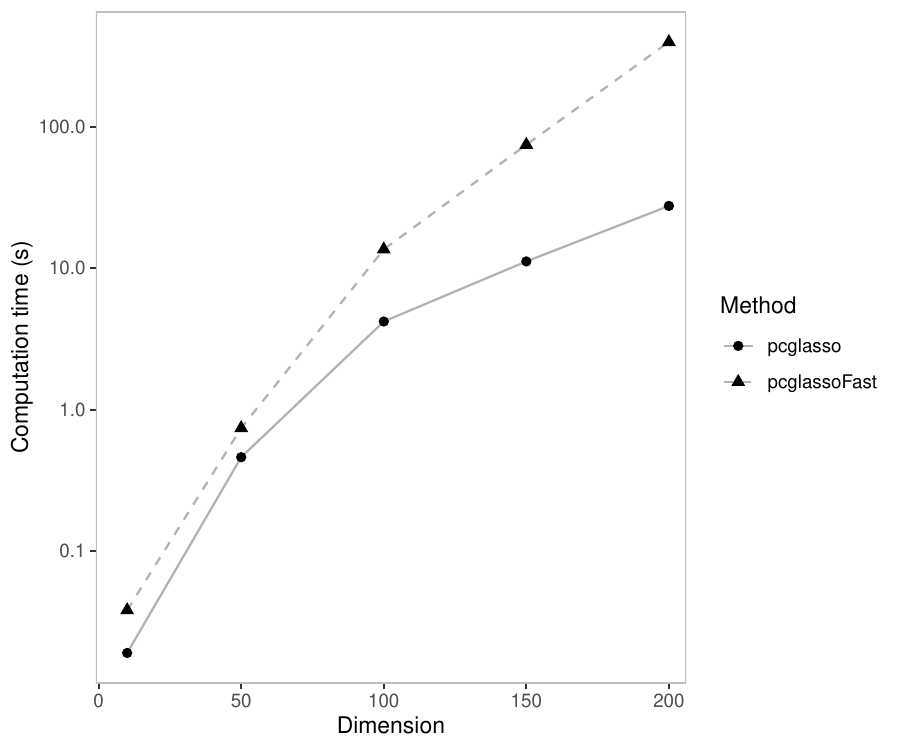} & \includegraphics[scale=0.55]{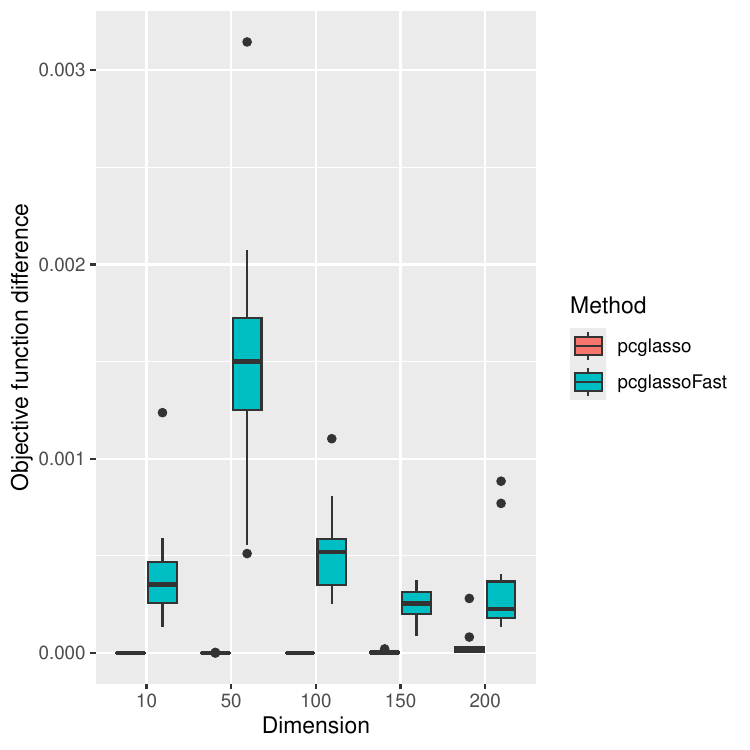} \\
    \includegraphics[scale=0.55]{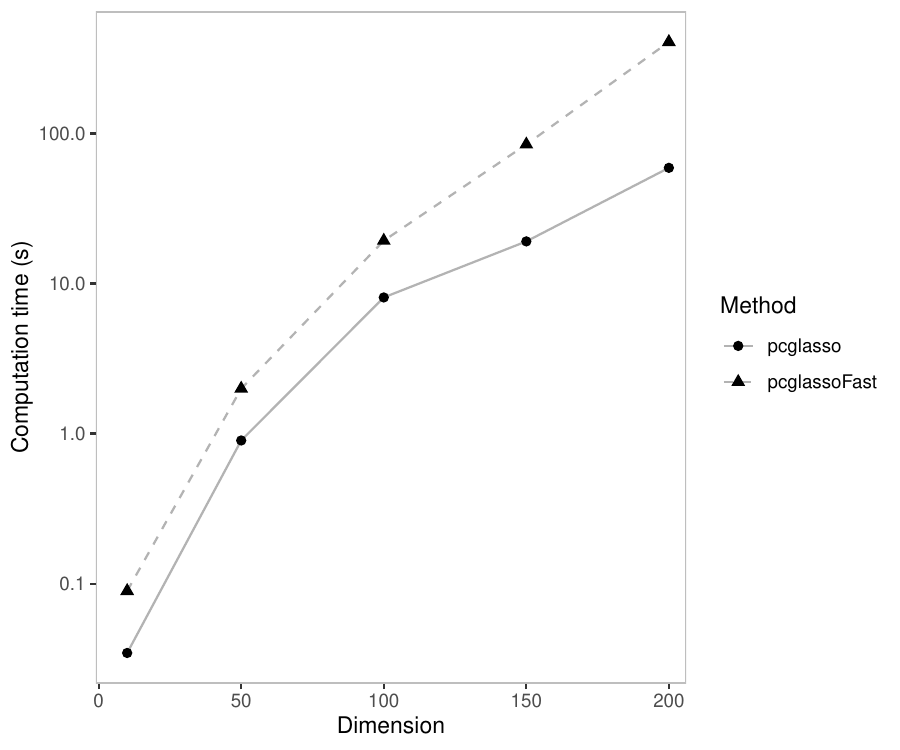} & \includegraphics[scale=0.55]{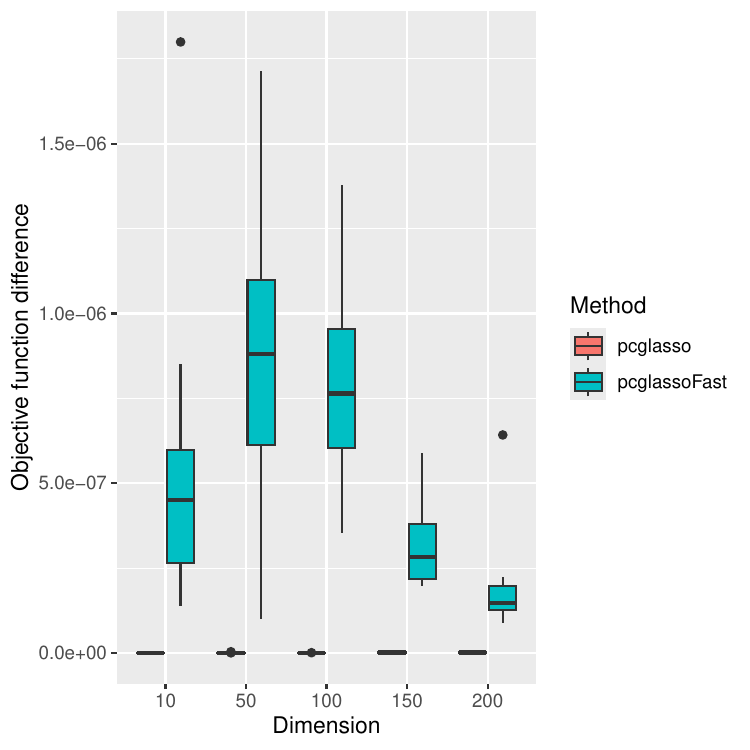}
\end{tabular}
\caption{Comparison of mean computation time (left) and distance to optimal value (right) for varying dimensions $p$ between pcglasso and pcglassoFast with the fast setting (top), balanced setting (middle) and convergence setting (bottom).}
\label{fig:FastDimension}
\end{figure}

\begin{figure}[hp]
\centering
\begin{tabular}{cc}
	\includegraphics[scale=0.55]{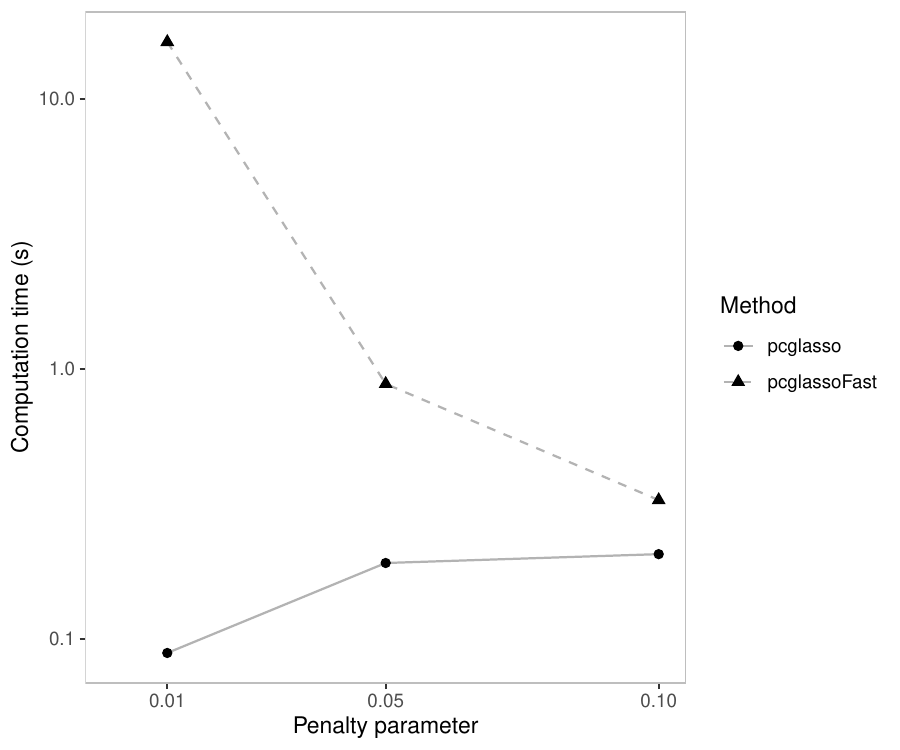} & \includegraphics[scale=0.55]{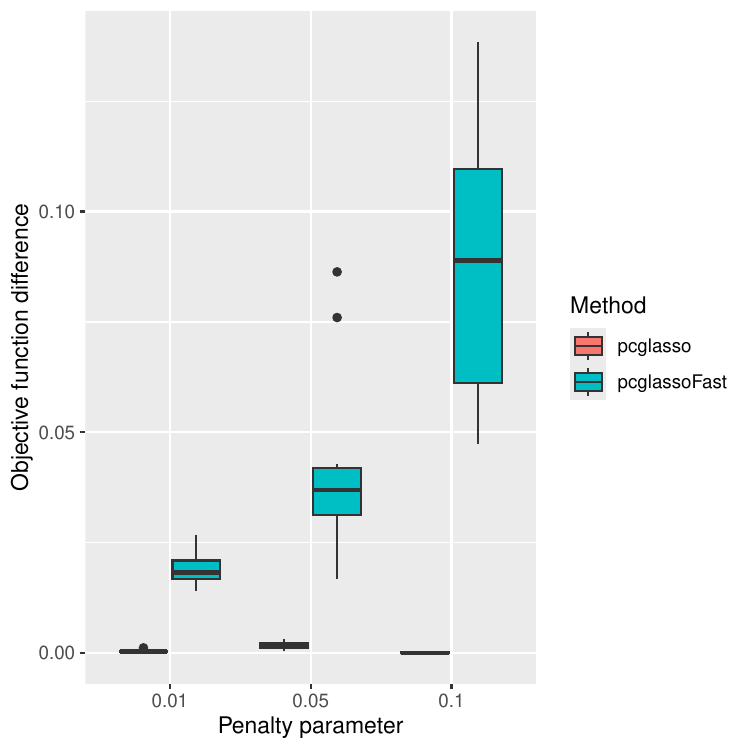} \\
	\includegraphics[scale=0.55]{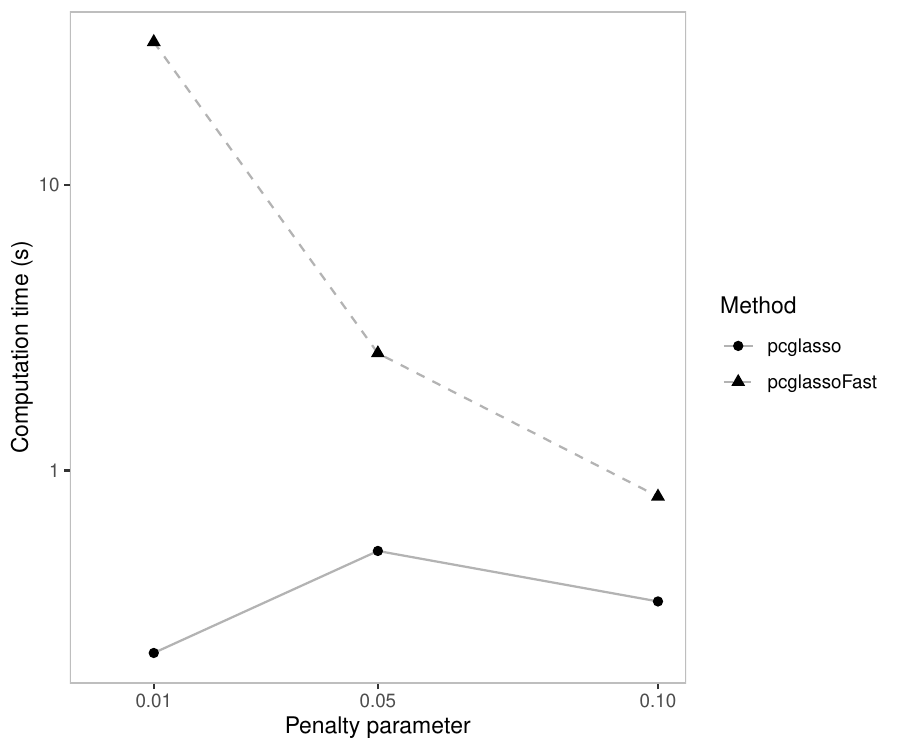} & \includegraphics[scale=0.55]{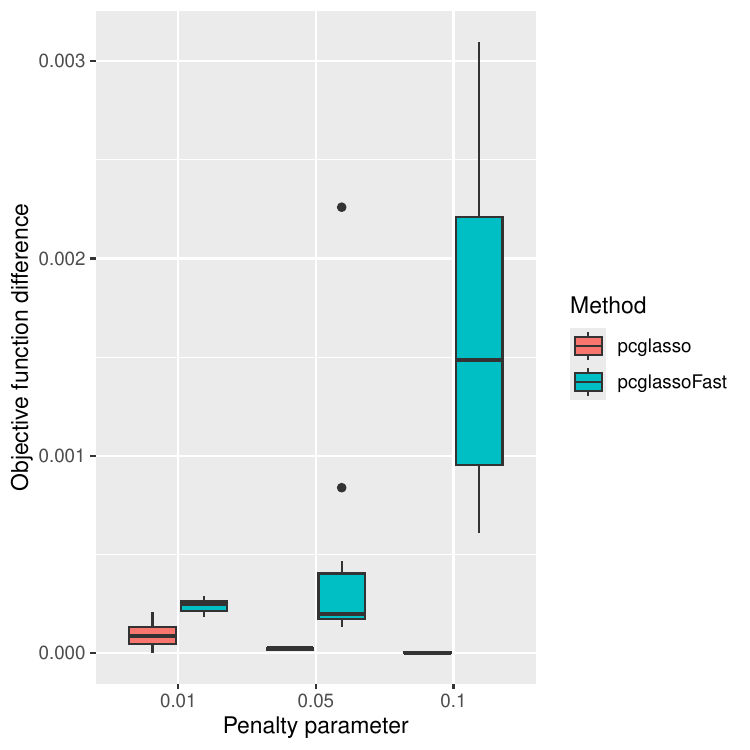} \\
    \includegraphics[scale=0.55]{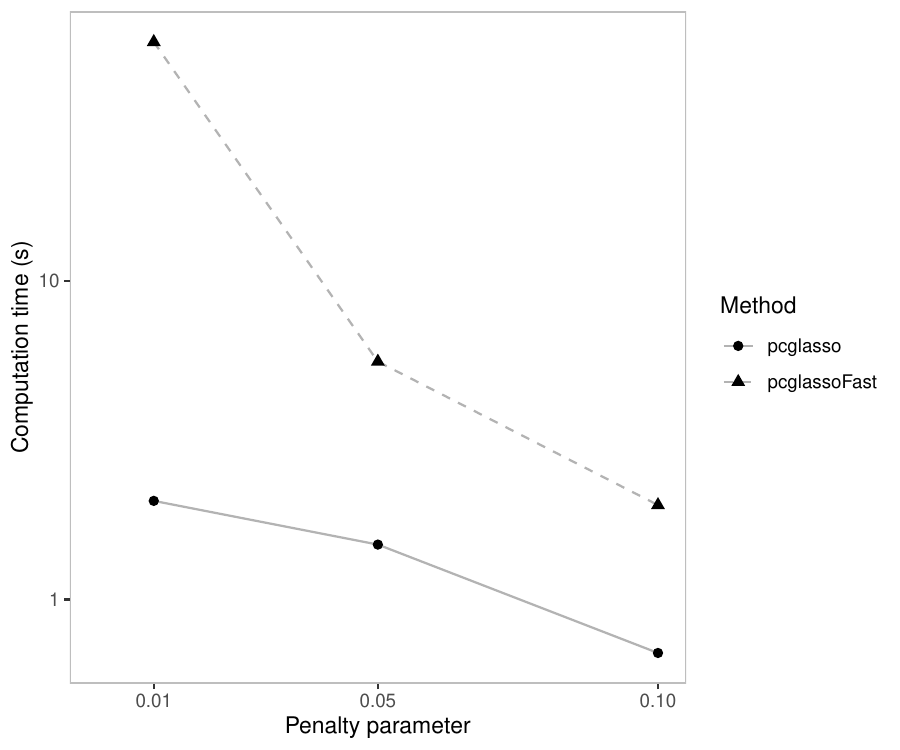} & \includegraphics[scale=0.55]{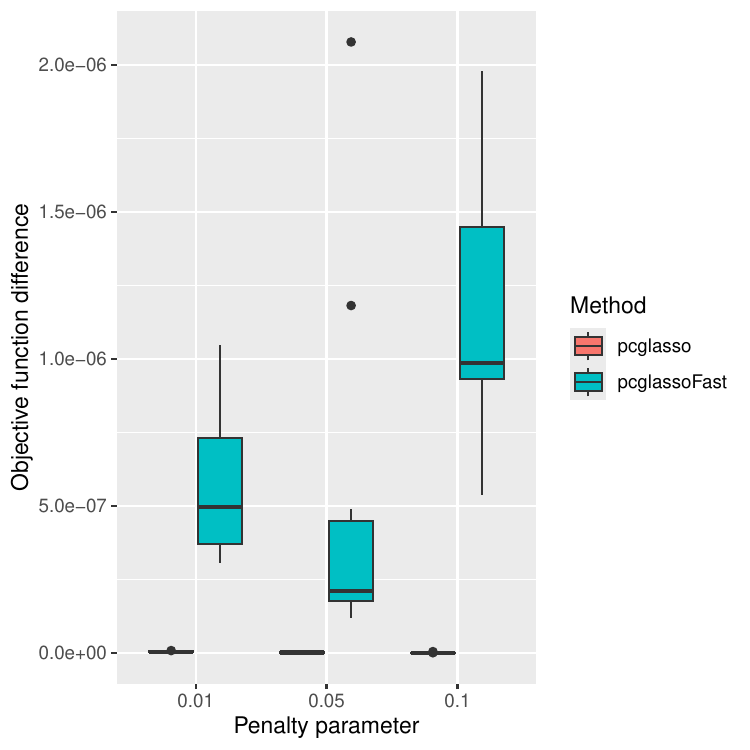}
\end{tabular}
\caption{Comparison of mean computation time (left) and distance to optimal value (right) for varying penalty parameter $\rho$ between pcglasso and pcglassoFast with the fast setting (top), balanced setting (middle) and convergence setting (bottom).}
\label{fig:FastPenalty}
\end{figure}

\begin{figure}[hp]
\centering
\begin{tabular}{cc}
	\includegraphics[scale=0.55]{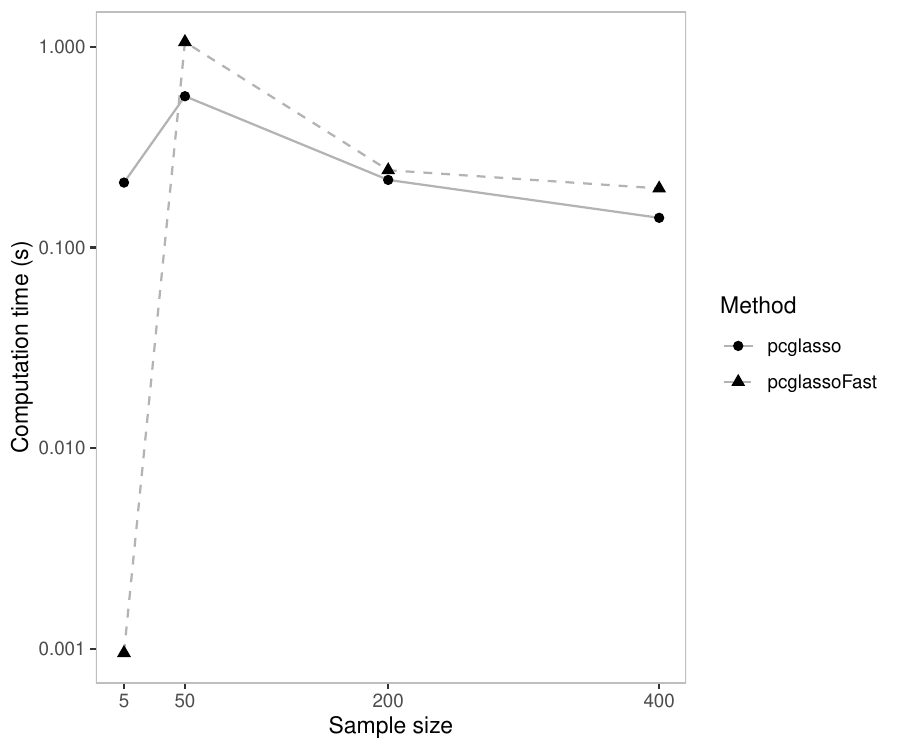} & \includegraphics[scale=0.55]{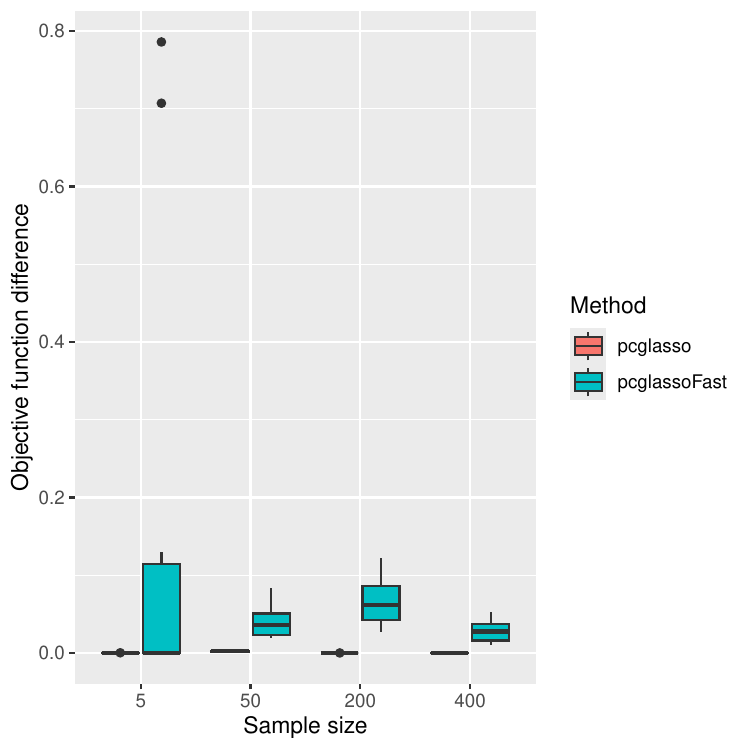} \\
	\includegraphics[scale=0.55]{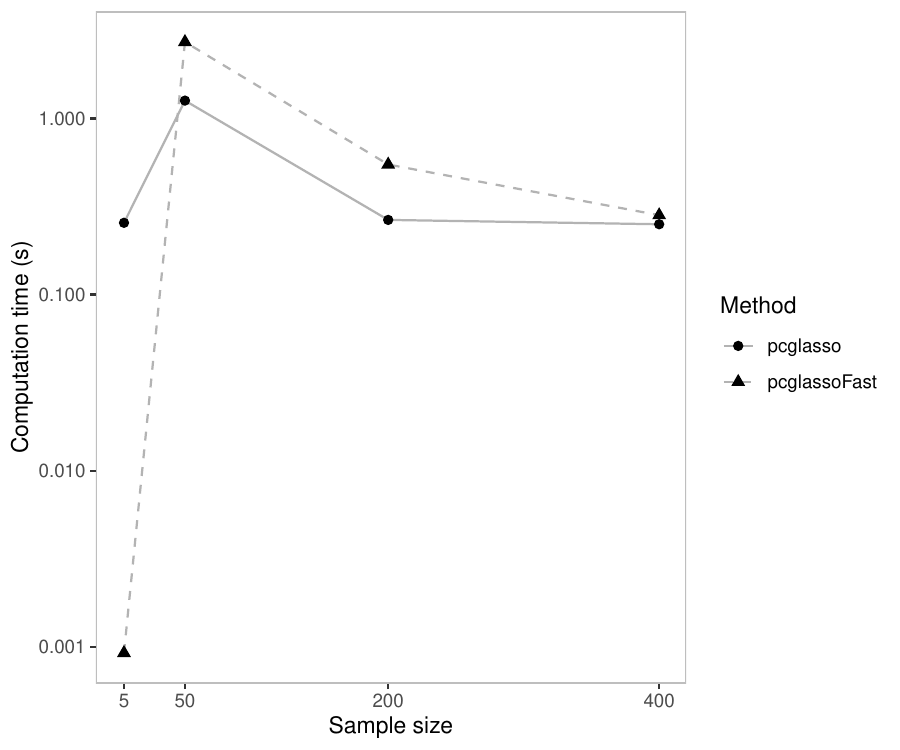} & \includegraphics[scale=0.55]{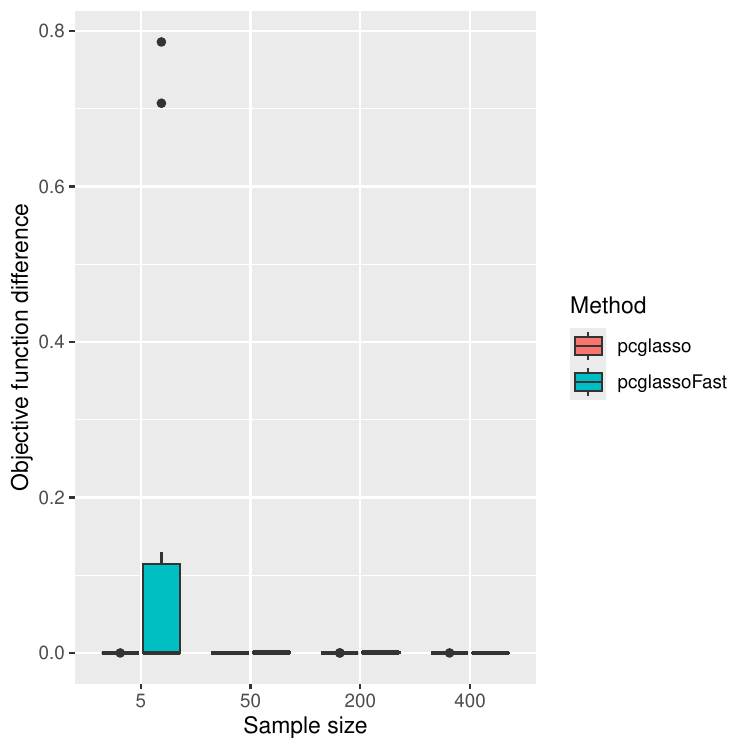} \\
    \includegraphics[scale=0.55]{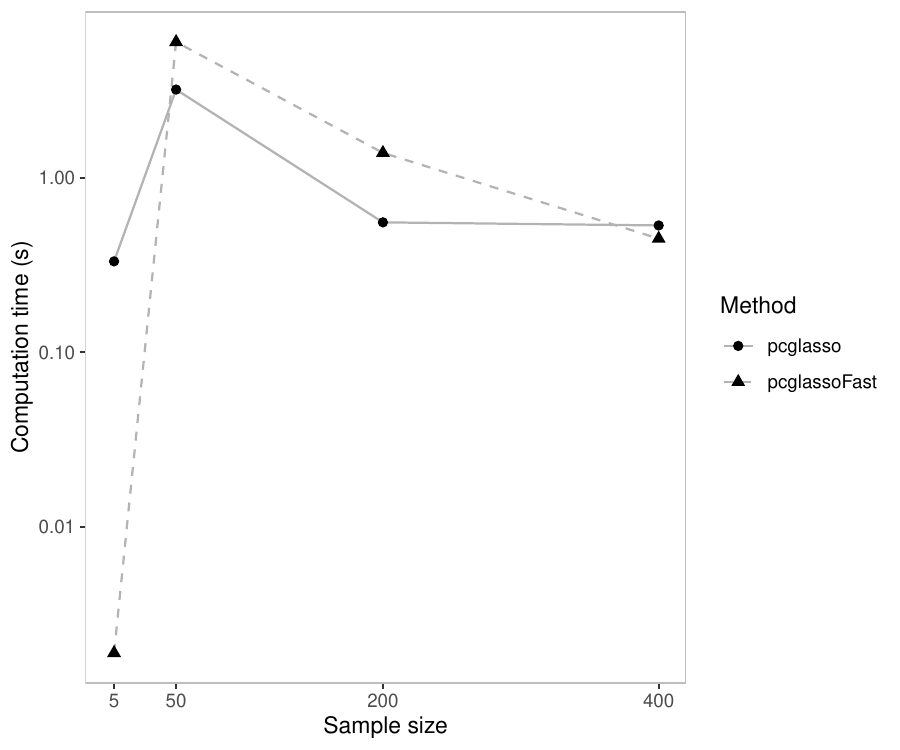} & \includegraphics[scale=0.55]{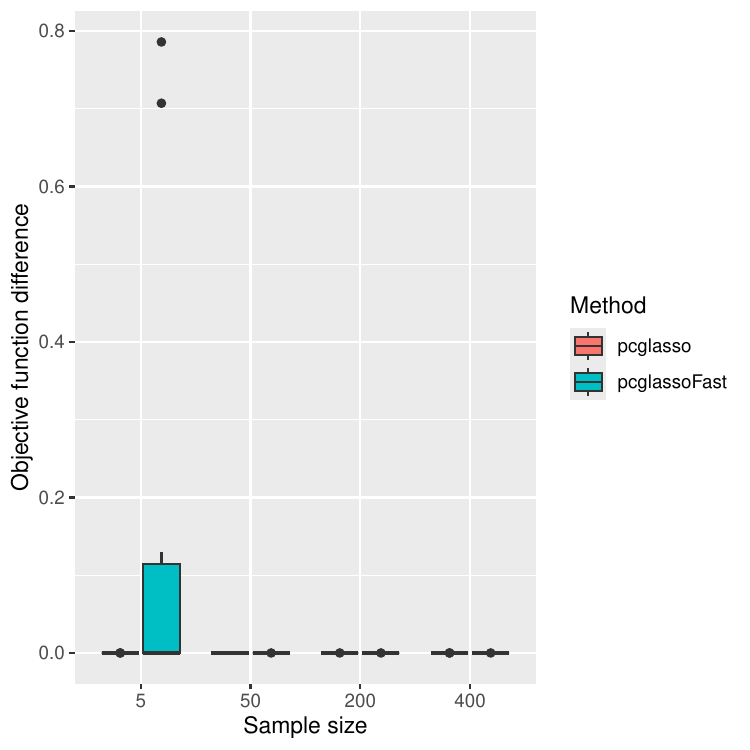}
\end{tabular}
\caption{Comparison of mean computation time (left) and distance to optimal value (right) for varying sample size $n$ between pcglasso and pcglassoFast with the fast setting (top), balanced setting (middle) and convergence setting (bottom).}
\label{fig:FastSampSize}
\end{figure}

Computation times for each of these settings along with the difference in objective function value when compared to the full convergence setting can be found in Figures \ref{fig:FastDimension}-\ref{fig:FastSampSize}. For different dimensions $p$ (Figure \ref{fig:FastDimension}), for each setting the pcglasso achieved smaller objective function values than pcglassoFast, and in almost all cases pcglasso had a lower mean computation time. This difference in mean computation time gets more pronounced in higher dimensions, suggesting that pcglasso enjoys better scaling. The only exception was in low dimensions for the fast setting were pcglassoFast had slightly faster mean computation time. However, this was at the expense of achieving a much worse objective function value.

For different penalty parameters $\rho$ (Figure \ref{fig:FastPenalty}), the pcglasso achieved smaller objective function values and faster mean computation time in all settings. Interestingly, pcglassoFast had significantly slower mean computation time for smaller penalty parameters, while the speed of pcglasso was less effected by the penalty parameter. The reason for this may be due to different starting points. For small penalty parameters, the optimum tends to be close to the inverse sample covariance matrix, the default starting point for pcglasso (when it exists). As the penalty parameter increases, the optimal partial correlation matrix shrinks towards the identity matrix, the default starting point for pcglassoFast. This suggests that our proposed starting value may be more suitable for path type implementations where estimates are obtained for a sequence of penalty parameters $\rho_1 < \dots < \rho_k$.

For different sample sizes $n$ (Figure \ref{fig:FastSampSize}), pcglasso also achieves lower objective function value and quicker mean computation for moderate sample sizes. For larger sample sizes the mean computation times become quite similar, although pcglasso does still arrive at a smaller objective function value. For the small $n=5$ sample size, the situation is more complicated. pcglassoFast has very fast computation for all threshold values. However, this is because it always returns $\Delta = I$. In some cases pcglasso also returns $\Delta = I$, but does so in a slower time and reaching a worse objective function value than pcglassoFast (due to minor differences in the returned $\xi$ values) - in fact in these cases pcglassoFast reaches a very slightly lower objective function value than the full convergence setting. However, in other cases pcglasso returns a non-diagonal $\Delta$ which achieves a lower objective function value. This is seen in the boxplots of Figure \ref{fig:FastSampSize} where the boxplot for pcglassoFast remains the same for all threshold values. It is possible that there is a local minimum at $\Delta = I$ which pcglassoFast is not able to escape, while pcglasso is able to instead reach a better optimum. Differing starting points may also explain this dynamic. Regardless of local vs global optima, it is usually more useful in practice to estimate non-diagonal $\Delta$, and so pcglasso's ability to find such optima is a good thing.

\section{Discussion}\label{sec:disc}

In this paper we have made significant advances in the computation of the PCGLASSO estimate for Gaussian graphical models.  The previously proposed coordinate descent method of \cite{Carter2023} was prohibitively slow and only worked for positive definite $S$ - when the sample size is larger than the dimension. In fact, it was not previously known if the PCGLASSO estimate even exists when $S$ is not positive definite. In this paper we not only proved that the PCGLASSO estimate does exist when $S$ is not positive definite - as long as it is generated from Gaussian data - and provided the range of values for $c$ for which it exists, but also proposed a computation method which works for non-positive definite $S$ and provides competitive performance in terms of speed. The algorithm is implemented in the \texttt{R} package \texttt{PCGLASSO} available at \url{https://github.com/JackStorrorCarter/PCGLASSO} and was the first publicly available implementation of the PCGLASSO.

While the implemented algorithm for the PCGLASSO is slower than available implementations of the GLASSO, this should be expected due to the increased complexity in the objective function and the advantages the PCGLASSO enjoys over the GLASSO in terms of estimation should justify this increased computation time. The implemented algorithm allows reasonable computation times for moderate dimensions - in the simulated examples it took approximately 30 seconds for problems of dimension $p=200$ in the star setting and only 2-4 seconds in the hub, AR2 and random graph settings. 

The pcglassoFast method of \cite{bogdan2025identifying} provides an interesting comparison for optimisation of the PCGLASSO. Similar to our proposed method, they use an alternating algorithm. However, in place of the DRS and FBS algorithms, an adapted version of the block coordinate descent algorithm of \cite{Banerjee2008} is used for the optimisation of $\Delta$ and a Newton method is used for optimisation of $\xi$. Furthermore, the optimisation of $\Delta$ uses an adapted version of the Fortran subroutine of \cite{Sustik2012}, while pcglasso is coded directly in R. This makes the results of \cite{bogdan2025identifying} Appendix 1, showing much faster computation times for pcglassoFast, quite believable. However, the results of this paper show this to be misleading because it did not consider the objective function value. With the current default settings, pcglassoFast is still quite far from the optimum when it terminates. When taking this into account, we have shown that pcglasso is actually currently faster than pcglassoFast.

Regardless of which method is faster, both pcglasso and pcglassoFast allow the PCGLASSO method to be implemented in a reasonable amount of time and show potential for even further improvement. The pcglasso implementation might be improved by using Fortran or C++ code. The pcglassoFast implementation may benefit from different default settings such as starting point and stopping rules, which were found to have a big effect in pcglasso. Since the two methods use different algorithms for both the $\Delta$ optimisation and $\xi$ optimisation, it may also be worth investigating different combinations of these to find the best pair.

\section*{Acknowledgements}
The research by JSC was supported in part by the MIUR Excellence Department Project awarded to Dipartimento di Matematica, Università di Genova, CUP D33C23001110001 and the 100021-BIPE 2020 grant and by the EUTOPIA Science and Innovation Fellowship Programme and funded by the European Union Horizon 2020 programme under the Marie Skłodowska-Curie grant agreement No 945380.

\medskip

\newpage
%\printbibliography
\bibliographystyle{plainnat}
\bibliography{myrefs}

\newpage

\appendix
\appendixpage

\section{Proof of Proposition \ref{prop}}\label{app:proof}

\begin{proof}
We will prove the solution existence with penalty parameter $\rho=0$. Since the penalty term $\rho \sum_{i \neq j} \vert \Delta_{ij} \vert$ is continuous and bounded, the solution existence follows for $\rho>0$.  We begin by rewriting the objective function in terms of the $\Theta$ parameterisation. Since $\log(\det(\Theta)) = \log(\det(\Delta)) + 2\sum_{i=1}^p \log(\xi_{ii})$ and $\xi_{ii} = \Theta_{ii}^{1/2}$, the objective function is
$$ -\log(\det(\Theta)) + (1 - c) \sum_{i=1}^p \log(\Theta_{ii}) + \mathrm{tr}(S\Theta) $$

The objective function can be further rewritten in terms of the eigenvalues and eigenvectors of $S$ and $\Theta$. Since $S$ and $\Theta$ are both symmetric, they are guaranteed to have an orthonormal basis of eigenvectors. The $p \times n$ matrix $X - \bar{X}$ is of rank $n-1$, so it follows from \cite[Theorem 8.3.3]{mathai2022multivariate} that $S$ has exactly $k = p - (n - 1)$ eigenvalues equal to $0$ while the remaining eigenvalues are strictly positive and distinct with probability 1 \citep{srivastava2003singular}. We write the eigenvalues of $S$ is ascending order $\lambda_1,\dots,\lambda_k=0$ and $0 < \lambda_{k+1} < \dots < \lambda_{p} < \infty$ with corresponding orthonormal eigenvectors $v_1,\dots,v_p$. The eigenvalues of positive definite $\Theta$ are $\sigma_1,\dots,\sigma_p > 0$ with corresponding orthonormal eigenvectors $w_1,\dots,w_p$ with $w_{ij}$ the $j$th entry of $w_i$.

The determinant is the product of the eigenvalues. Using the eigendecompositions of $\Theta$ and $S$, the diagonal entries of $\Theta$ can be written as $\Theta_{jj} = \sum_{i=1}^p \sigma_i w_{ij}^2$, while the trace term can be written as $\mathrm{tr}(S\Theta) = \sum_{i,j=1}^{p} \sigma_i \lambda_j ( w_i^{\T} v_{j} )^2$. Hence, the objective function in terms of eigenvalues and eigenvectors is
$$ -\sum_{i=1}^p \log( \sigma_i ) + (1 - c) \sum_{j=1}^p \log\left( \sum_{i=1}^p \sigma_i w_{ij}^2 \right) + \sum_{i,j=1}^{p} \sigma_i \lambda_j ( w_i^{\T} v_{j} )^2 $$

We will show that the objective function tends to $\infty$ as $\Theta$ tends towards the boundary of the feasible region. Because the objective function is continuous, this is enough to prove the solution existence. The feasible region is $\Theta \in \mathcal{S}$, the set of positive definite matrices, which is characterised by $\sigma_1,\dots,\sigma_p > 0$ and so the boundary of the space is when $\sigma_i \rightarrow 0$ or $\sigma_i \rightarrow \infty$. The set of possible orthonormal eigenvectors of $\Theta$ is closed and so we show that the objective function tends to $\infty$ as any $\sigma_i \rightarrow 0$ or $\sigma_i \rightarrow \infty$ for any fixed eigenvectors.

Fix the eigenvectors $w_1,\dots,w_p$ and suppose that $w_1,\dots,w_l$ are in the null space of $S$. This means that $w_1,\dots,w_l$ are each orthogonal to all of $v_{k+1},\dots,v_p$, while $w_{l+1},\dots,w_p$ are not orthogonal to at least one of $v_{k+1},\dots,v_p$. We must have $l \leq k$, otherwise we have more than $p$ orthogonal vectors of length $p$.

First assume that $w_1,\dots,w_l$ have no entries equal to zero. Suppose that we allow some eigenvalues to tend to $0$, $\sigma_i \rightarrow 0$ for $i \in I_0$, some eigenvalues to tend to $\infty$, $\sigma_i \rightarrow \infty$ for $i \in I_\infty$, while the other eigenvalues remain finite and away from $0$. We denote $I_{\infty,l} = I_{\infty} \cap \{ 1,\dots,l \}$ those indices for which $\sigma_i \rightarrow \infty$ and $w_i$ is in the null space of $S$, and let $m = \abs{I_{\infty,l}}$. 

If $I_0 = \{ 1,\dots,p \}$ so that $\sigma_i \rightarrow 0$ for all $i$, then the objective function tends to $\infty$ because $1-c < 1$. Otherwise, if $m=0$, when $\sigma_i \rightarrow \infty$, the trace term tends to $\infty$ at a linear rate since $w_i$ is not in the null space of $S$, while as $\sigma_j \rightarrow 0$ the first logarithmic term tends to $\infty$ and the other terms remain finite. Hence the objective function tends to $\infty$. If $m>0$, considering the second term in the objective function, we obtain the following bound
\begin{align}\label{eq:inequality}
    \sum_{j=1}^p \log\left( \sum_{i=1}^p \sigma_i w_{ij}^2 \right) &= \log\left( \prod_{j=1}^p \sum_{i=1}^p \sigma_i w_{ij}^2 \right) \nonumber \\
    &\geq \log\left( \sum_{i \in I_{\infty,l} } \sigma_i^p \prod_{j=1}^p w_{ij}^2 \right) \nonumber \\
    &\geq \log(m) + \frac{1}{m} \sum_{i \in I_{\infty,l}} \log\left( \sigma_i^p \prod_{j=1}^p w_{ij}^2 \right) \\
    &= \log(m) + \frac{p}{m} \sum_{i \in I_{\infty,l}} \log\left( \sigma_i \right) + \frac{1}{m} \sum_{i \in I_{\infty,l}} \sum_{j=1}^p \log( w_{ij}^2 ) \nonumber \\
    &= \frac{p}{m} \sum_{i \in I_{\infty,l}} \log\left( \sigma_i \right) + \textrm{const.} \nonumber
\end{align}
The first inequality follows by noting that $\prod_{j=1}^p \sum_{i=1}^p \sigma_i w_{ij}^2$ is a polynomial in the $\sigma_i$ with positive coefficients. Since $\sigma_i > 0$, removing any term of the polynomial makes it smaller. We remove all terms except the $\sigma_i^p$ terms for $i \in I_{\infty,l}$. The second inequality is a result of Jensen's inequality. The final line includes the term $\textrm{const.}$ which does not depend on the eigenvalues of $\Theta$.

Using this inequality we obtain the following lower bound for the objective function
\begin{align*}
-\sum_{i=1}^p \log( \sigma_i ) &+ (1 - c) \sum_{j=1}^p \log\left( \sum_{i=1}^p \sigma_i w_{ij}^2 \right) + \sum_{i,j=1}^{p} \sigma_i \lambda_j ( w_i^{\T} v_{j} )^2 \\
& \geq -\sum_{i=1}^p \log( \sigma_i ) + (1 - c) \frac{p}{m} \sum_{i \in I_{\infty,l}} \log\left( \sigma_i \right) + \sum_{i,j=1}^{p} \sigma_i \lambda_j ( w_i^{\T} v_{j} )^2 + (1-c) \mathrm{const.} \\
& = \left( \frac{(1-c)p}{m} - 1 \right) \sum_{i \in I_{\infty,l}} \log( \sigma_i ) - \sum_{i \not\in I_{\infty,l}} \log(\sigma_i) + \sum_{i=l+1}^p \sigma_i \sum_{j=k+1}^{p} \lambda_j ( w_i^{\T} v_{j} )^2 + (1-c) \mathrm{const.}
\end{align*}

In the term with the double sum, the second sum begins at $j=k+1$ because $\lambda_1,\dots,\lambda_k=0$, while the first sum begins at $i=l+1$ because $w_1,\dots,w_l$ are in the null space of $S$ so $w_i^\T v_j = 0$ for $i=1,\dots,l$, $j=k+1,\dots,p$. However, for all $i=l+1,\dots,p$, because $w_i$ is not in the null space of $S$, there is a $j \in \{ k+1,\dots,p \}$ such that $w_i^\T v_j \neq 0$. Hence $\sum_{j=k+1}^{p} \lambda_j ( w_i^{\T} v_{j} )^2 > 0$ for all $i=l+1,\dots,p$.

This lower bound separates in the eigenvalues and so we can analyse the contribution of each eigenvalue individually.
For $i \in I_{\infty,l}$, the contribution is $\left( \frac{(1-c)p}{m} - 1 \right) \log( \sigma_i )$ and this term tends to $\infty$ as $\sigma_i \rightarrow \infty$ as long as $\frac{(1-c)p}{m} - 1 > 0$, which occurs when $c < 1 - \frac{m}{p}$. 
For other $i \in I_\infty$ (but not in $I_{\infty,l}$), the contribution is $-\log(\sigma_i) + \sigma_i \sum_{j=k+1}^{p} \lambda_j ( w_i^{\T} v_{j} )^2$ and this tends to $\infty$ as $\sigma_i \rightarrow \infty$ since $\sum_{j=k+1}^{p} \lambda_j ( w_i^{\T} v_{j} )^2 > 0$. 
For $i \in I_0$, the contribution is $-\log(\sigma_i) + \sigma_i \sum_{j=k+1}^{p} \lambda_j ( w_i^{\T} v_{j} )^2$, and this tends to $\infty$ as $\sigma_i \rightarrow 0$. 
Finally, for all other $i$ (not in $I_\infty$ or $I_0$), the contribution is also $-\log(\sigma_i) + \sigma_i \sum_{j=k+1}^{p} \lambda_j ( w_i^{\T} v_{j} )^2$, and this is finite valued for any finite $\sigma_i > 0$. 
Hence the lower bound, and therefore the objective function, tends to $\infty$ as $\sigma_i \rightarrow \infty$, $i \in I_{\infty}$ and $\sigma_i \rightarrow 0$, $i \in I_0$, as long as $c < 1 - \frac{m}{p}$. This remains valid for any choice of $I_\infty$ and $I_0$. The bound on $c$ is most tight when $m$ is taken to be as large as possible. This occurs when $m=l=k$, making the bound $c < 1 - \frac{k}{p}$.

So far it has been assumed that the eigenvectors of $\Theta$ contain no zero entries. This was important because when there are zeros in the eigenvectors $w_1,\dots,w_l$, the inequality (\ref{eq:inequality}) can have $\log(0)$ terms and the argument breaks down. To relax this assumption, again suppose that $w_1,\dots,w_l$ are in the null space of $S$ while $w_{l+1},\dots,w_p$ are not in the null space of $S$. Now the eigenvectors $w_{l+1},\dots,w_p$ are allowed to have zeros in any combination, while zeros in $w_1,\dots,w_l$ are only allowed to appear in specific ways: 
%each $w_i$ can contain at most $k-1$ zeros; if both $w_i$ and $w_j$ have $k-1$ zeros, then the positions of the zeros cannot be all the same; if three $w_i$ have at least $k-2$ zeros, then they cannot all share the same position of the $k-2$ zeros etc. 
each $w_i$ can contain at most $k-1$ zeros. If $w_i$ has $k-1$ zeros then all other $w_j$ can have at most $k-2$ zeros. If $w_i$ has $k-1$ zeros and $w_j$ has $k-2$ zeros then all other $w_h$ can have at most $k-3$ zeros etc.

Using these properties, one is able to choose terms from the polynomial $\prod_{j=1}^p \sum_{i=1}^p \sigma_i w_{ij}^2$ with non-zero coefficients to create an inequality similar to that in (\ref{eq:inequality}). For example, when $l=1$ and $I_{\infty,l} = \{ 1 \}$, this polynomial contains a term $a \sigma_1^{p-(k-1)} \sigma_{i_2} \dots \sigma_{i_k}$ with distinct $i_2,\dots,i_k \neq 1$ where $$a = w_{1j_1}^2 \dots w_{1j_{p-(k-1)}}^2 w_{i_2j_{p-(k-2)}}^2 \dots w_{i_kj_p}^2 \neq 0$$ with distinct $j_1,\dots,j_p$. Removing all except this term from the polynomial gives the inequality
\begin{align}
    \log\left( \prod_{j=1}^p \sum_{i=1}^p \sigma_i w_{ij}^2 \right)
    &\geq \log\left( a \sigma_1^{p-(k-1)} \sigma_{i_2} \dots \sigma_{i_k} \right) \nonumber \\
    &= (p-(k-1)) \log(\sigma_1) + \sum_{j=2}^k \log\left( \sigma_{i_j} \right) + \log(a) \nonumber
\end{align}
Using this to lower bound the objective function, 
%\begin{align*}
%&\sum_{i=1}^p \log( \sigma_i ) - (1 - c) \sum_{j=1}^p \log\left( \sum_{i=1}^p \sigma_i w_{ij}^2 \right) - \sum_{i,j=1}^{p} \sigma_i \lambda_j ( w_i^{\T} v_{j} )^2 \\
%& \leq \left( 1-(1-c)(p-(k-1)) \right) \log( \sigma_1 ) + \sum_{i=2}^p \log(\sigma_i) - \sum_{i,j=1}^{p} \sigma_i \lambda_j ( w_i^{\T} v_{j} )^2 - (1-c) \left( \sum_{j=2}^k \log(\sigma_{i_j}) + \log(a) \right)
%\end{align*}
we find that the objective function tends to $\infty$ as $\sigma_1 \rightarrow \infty$ when $c < 1 - \frac{1}{p-(k-1)}$, which is less strict than $c < 1 - \frac{k}{p}$. Note also that this lower bound still tends to $\infty$ as $\sigma_i \rightarrow 0$ for $i \neq 1$ because the $\sigma_{i_j}$ are distinct.

In the other extreme, when $l=k$ and $I_{\infty,l} = \{ 1,\dots,l \}$, this polynomial contains a term $a_1 \sigma_1^{p-(k-1)} \sigma_2 \dots \sigma_k$ where $a_1 = w_{1j_1}^2 \dots w_{1j_{p-(k-1)}}^2 w_{2j_{p-(k-2)}}^2 \dots w_{kj_p}^2$ with all $w_{ij} \neq 0$ and distinct $j_1,\dots,j_p$. Similarly, it contains terms $a_2 \sigma_1 \sigma_2^{p-(k-1)} \dots \sigma_k,\dots,a_k \sigma_1 \sigma_2 \dots \sigma_k^{p-(k-1)}$. Hence we have the following bound
\begin{align}
    \log\left( \prod_{j=1}^p \sum_{i=1}^p \sigma_i w_{ij}^2 \right)
    &\geq \log\left( a_1 \sigma_1^{p-(k-1)} \sigma_2 \dots \sigma_k + \dots + a_k \sigma_1 \sigma_2 \dots \sigma_k^{p-(k-1)} \right) \nonumber \\
    &\geq \log(k) + \frac{1}{k} \left( \log\left(a_1 \sigma_1^{p-(k-1)} \sigma_2 \dots \sigma_k\right) + \dots + \log\left(a_k \sigma_1 \sigma_2 \dots \sigma_k^{p-(k-1)}\right) \right) \nonumber \\
    &= \log(k) + \frac{1}{k} \left( \sum_{i=1}^k \log\left( \sigma_i^p \right) + \log(a_i) \right) \nonumber \\
    &= \frac{p}{k} \sum_{i=1}^k \log\left( \sigma_i \right) + \textrm{const.} \nonumber
\end{align}
where the second inequality is again a result of Jensen's inequality. The lower bound on the objective function is then as in the original case.

For general $l$, the same strategy can be used as above, choosing terms $a_1 \sigma_1^{p-(k-1)} \sigma_2 \dots \sigma_l \sigma_{i_{1,l+1}} \dots \sigma_{,i_{1,k}}, \dots,$ $a_l \sigma_1 \dots \sigma_l^{p-(k-1)} \sigma_{i_{l,l+1}} \dots \sigma_{i_{l,k}}$ where the $i_{j,l+1},\dots,i_{j,k} \neq 1,\dots,l$ are distinct, to find that the lower bound tends to $\infty$ as long as $c < 1 - \frac{l}{p-(k-l)}$, which is less strict than $c < 1 - \frac{k}{p}$.

In the above it was assumed that the eigenvectors $w_1,\dots,w_l$ in the null space of $S$ can only have zeros appearing in specific ways. Since $S$ is a Gaussian sample covariance matrix, it has singular Wishart distribution and so with probability 1 its non-zero eigenvalues are distinct and the corresponding eigenvectors are continuously distributed. Hence the null space of $S$ is a continuously distributed, $k$ dimensional linear subspace of $\mathbb{R}^p$. It follows that the null space intersects non-trivially (i.e. discounting the zero vector) with any subspace of dimension up to $p-k$ with probability 0.

The set of $p$ dimensional vectors with $k$ (or more) zeros is a finite union of $p-k$ (or smaller) dimensional subspaces and so intersects with the null space with probability 0. For any fixed non-zero vector $w$ in the null space, the set of vectors in the null space that are orthogonal to $w$ is a subspace of $\mathbb{R}^p$ of dimension $k-1$. Hence, by the same logic as above, this new subspace contains no vectors with $k-1$ or more zeros with probability 1. If follows that the null space contains no two orthogonal vectors each with $k-1$ zeros with probability 1. Continuing this logic, it follows that the given conditions about the number of zero entries in $w_1,\dots,w_l$ occur with probability 1 when $S$ is a Gaussian sample covariance matrix.

This shows that with probability 1, the continuous objective function tends towards $\infty$ as eigenvalues of $\Theta$ tend towards either $0$ or $\infty$, implying that an optimum exists for $c < 1 - k/p$.
\end{proof}

\section{Proximal point operators}\label{app:prox}

Here we derive the proximal point operators for the three functions used in the the DRS algorithm (Section \ref{subsec:DR}) and the FBS algorithm (Section \ref{subsec:FB}). Recall that these functions are
\begin{equation*}
    f(\Delta) = -\log \det (\Delta) + \iota_{\calS}(\Delta)
\end{equation*}
where $\calS$ is the set of positive definite, $p \times p$ matrices,
\begin{equation*}
    g(\Delta) = \rho \sum_{i\neq j} \abs{\Delta_{ij}} + \mathrm{tr}\left( \tilde{S} \Delta \right) + \iota_{M_1}(\Delta)
\end{equation*}
where $M_1$ is the set of symmetric $p \times p$ matrices with unit diagonal, and
\begin{equation*}
    h(\xi) = -2c\sum_{i=1}^p \log(\xi_{i}) + \iota_{\R^p_+}(\xi_1,\dots,\xi_p)
\end{equation*}
and that the proximal point operator for a function $a:\mathcal{X} \rightarrow [-\infty,\infty]$ is 
$$\prox_{a}(v) = \argmin_{x \in \mathcal{X}}\left( a(x) + \frac{1}{2} \norm{x - v}{\mathcal{X}}^2\right) $$

The function $f$ is convex, is only finite for positive definite $\Delta$ and only depends on the eigenvalues of $\Delta$. By the Spectral Theorem, any symmetric matrix can be written as $\Delta = V \Sigma V^\T$ where $\Sigma = \diag(\sigma)$, $\sigma = (\sigma_1,\dots,\sigma_p)$ are the eigenvalues of $\Delta$ and $V$ is the matrix with columns equal to the eigenvectors of $\Delta$. Then we can rewrite $f$ as
\begin{align*}
  f(\Delta) &= -\log\left(\prod_{i=1}^p \sigma_i \right) + \iota_{\R^p_+}(\sigma) \\
  & := \tilde{f}(\sigma)  
\end{align*}
and by \cite[Corollary 24.65]{BauComNew}, the proximal point operator satisfies
$$\prox_{\alpha f}(\Delta) = V \, \diag\left(\prox_{\alpha \tilde{f}}(\sigma)\right) \, V^\T$$
where 
\begin{align*}
        \prox_{\alpha \tilde{f}}(\sigma) & = \argmin_{\lambda\in \R^p_+}\left( -\log\left(\prod_{i=1}^p \lambda_i \right) + \frac{1}{2\alpha}\norm{\lambda-\sigma}{}^2\right) \\
        & = \argmin_{\lambda\in \R^p_+}\left( -\sum_{i=1}^p \log(\lambda_i) + \frac{1}{2\alpha} \sum_{i=1}^p (\lambda_i-\sigma_i)^2\right)
\end{align*}
This is separable by components and so
\begin{align*}
        \left( \prox_{\alpha \tilde{f}}(\sigma) \right)_i
        & = \argmin_{\lambda_i \in \R_+}\left( -\log(\lambda_i) + \frac{1}{2\alpha}\left(\lambda_i-\sigma_i\right)^2\right) \\
        &= \frac{1}{2}\left(\sigma_i+\sqrt{\sigma_i^2+4\alpha}\right)
\end{align*}

For $g$, the proximal point operator is
\begin{align*}
        \prox_{\alpha g}(\Delta) & = \argmin_{\Psi \in M_1}\left( \rho \sum_{i\neq j} \abs{\psi_{ij}} + \mathrm{tr}\left( \tilde{S} \Psi \right) +\frac{1}{2\alpha}\norm{\Psi-\Delta}{}^2\right)\\
        & = \argmin_{\Psi\in M_1}\left( \rho \sum_{i\neq j} \abs{\psi_{ij}} + \sum_{i,j=1}^p \tilde{S}_{ij}\psi_{ij} + \frac{1}{2\alpha}\sum_{i,j=1}^p(\psi_{ij}-\Delta_{ij})^2\right)
\end{align*}
This is again separable by components. The diagonal entries are equal to 1 due to the $M_1$ constraint
\begin{equation*}
        \left(\prox_{\alpha g}(\Delta)\right)_{ii} = 1
\end{equation*}
For $i \neq j$, the off diagonals are
\begin{align*}
        \left(\prox_{\alpha g}(\Delta)\right)_{ij} & = \argmin_{\psi \in \R}\left( \rho \abs{\psi} + \tilde{S}_{ij}\psi + \frac{1}{2\alpha}(\psi-\Delta_{ij})^2\right)\\
        & = \argmin_{\psi \in \R} \left( \abs{\psi} + \frac{1}{2\alpha\rho}\left(\psi-(\Delta_{ij}-\alpha \tilde{S}_{ij})\right)^2 \right)\\
        & = \mathrm{shrink}(\Delta_{ij}-\alpha \tilde{S}_{ij},\alpha \rho).
\end{align*}

For $h$ the proximal point operator is
\begin{equation*}
    \prox_{\gamma h}(\xi) = \argmin_{(t_1,\dots,t_p)\in\R_+^p}\left( -2c \sum_{i=1}^p \log(t_i) + \frac{1}{2\gamma}\sum_{i=1}^p(t_i-\xi_i)^2 \right)
\end{equation*}
which is separable by components. The $i$th entry is
\begin{align*}
    \left(\prox_{\gamma h}(\xi)\right)_i &= \argmin_{t_i\in\R_+} \left( -2c \log(t_i) + \frac{1}{2\gamma}(t_i-\xi_i)^2 \right) \\
    &= \frac{1}{2}\left(\xi_i+\sqrt{\xi_i^2+8c\gamma}\right)
\end{align*}

\section{Further testing}\label{app:furthertesting}

The testing in Section \ref{sec:test} was repeated for three different data generating $\Theta$: a hub graph, AR2 model and random graph. Each of these $\Theta$ have diagonal entries $\theta_{ii}=1$ and off-diagonal entries defined as

\begin{itemize}
\item Hub graph - partition the $p$ variables into groups of 5, with each group associated to a `hub' variable $i$.  For any $j\neq i$ in the same group as $i$ we set $\theta_{ij} = \theta_{ji} = \frac{-2}{\sqrt{p}}$ and otherwise $\theta_{ij}=0$.

\item AR2 model - $ \theta_{ij} = \begin{cases}
\frac{1}{2}, & j = i-1,i+1 \\
\frac{1}{4}, & j = i-2,i+2 \\
0, & \text{otherwise}
\end{cases} $

\item Random graph - randomly select $\frac{3}{2}p$ of the $\theta_{ij}$ and set their values to be uniform on $[-1,-0.4] \cup [0.4,1]$, and the remaining $\theta_{ij}=0$. Calculate the sum of absolute values of off-diagonal entries for each column.  Divide each off-diagonal entry by 1.1 times the corresponding column sum and average this rescaled matrix with its transpose to obtain a symmetric, positive definite matrix.
\end{itemize}

As in Section \ref{sec:test}, the proposed algorithm for the PCGLASSO is compared to a coordinate descent algorithm for the PCGLASSO and well as publicly available implementations of the GLASSO and SCAD penalised likelihoods. The results and conclusions closely match those outlined in Section \ref{sec:test} for the star graph, but with generally much quicker computation times than the star setting.

\begin{figure}[h]
\centering
\begin{tabular}{ccc}
   \includegraphics[scale=0.375]{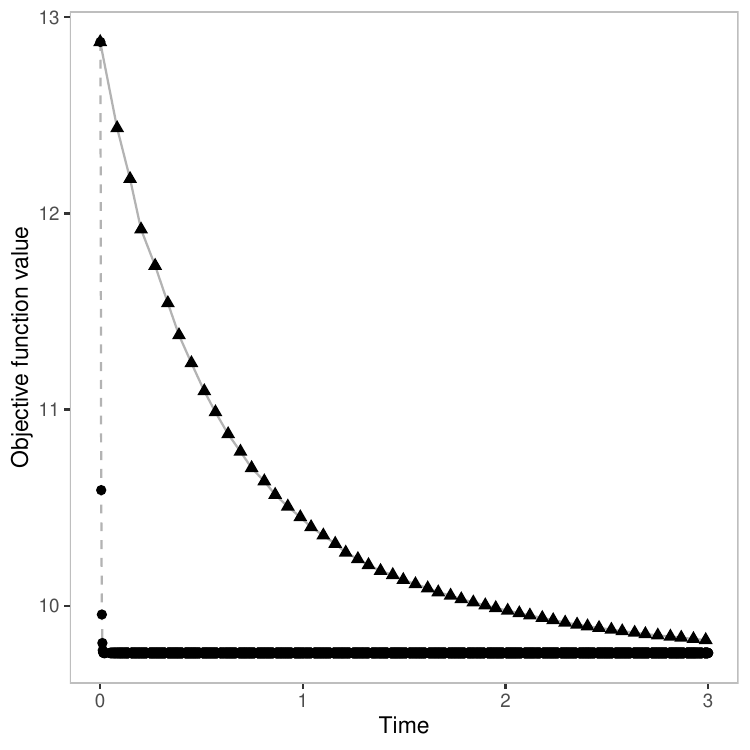}  &  \includegraphics[scale=0.375]{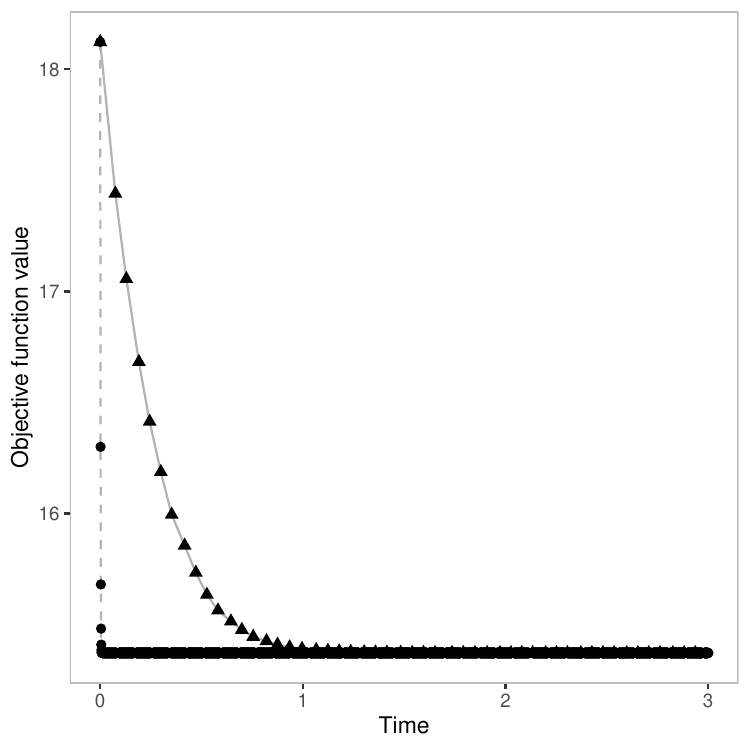}  & \includegraphics[scale=0.375]{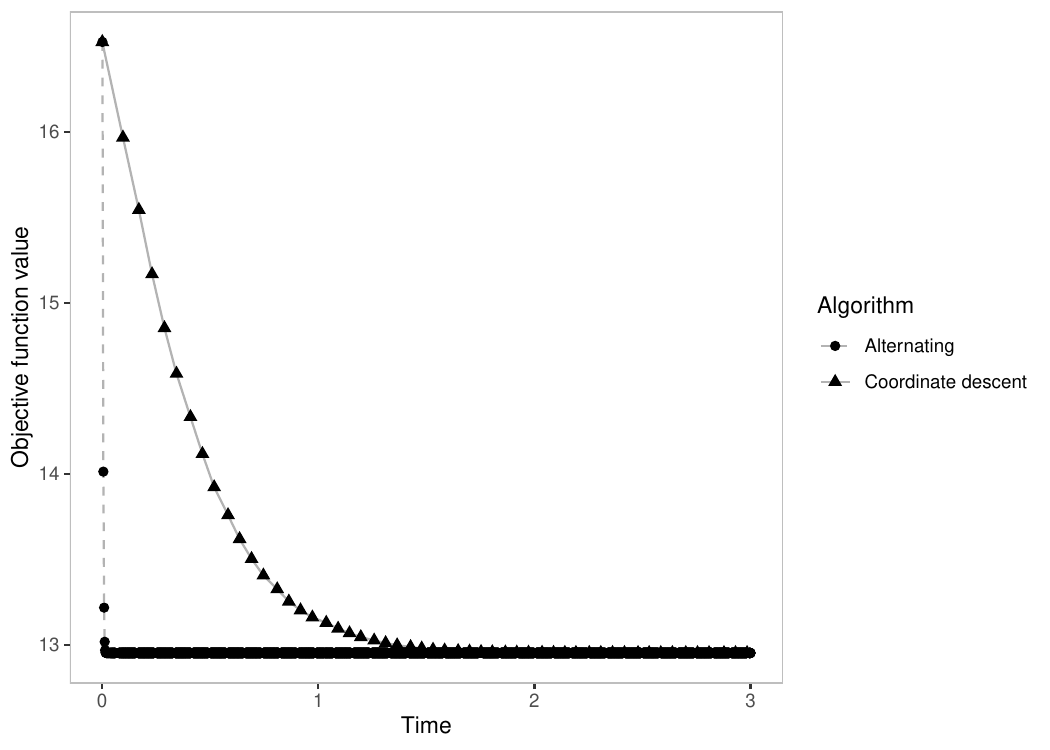}
\end{tabular}
\caption{Comparison of the proposed alternating algorithm to a coordinate descent algorithm for the hub graph (left), AR2 model (centre) and random graph (right) with $p=20$, $n=40$. Points correspond to each iteration of the algorithms.}
\label{fig:CD}
\end{figure}

\begin{figure}[hp]
\centering
\begin{tabular}{c}
	\includegraphics[scale=0.6]{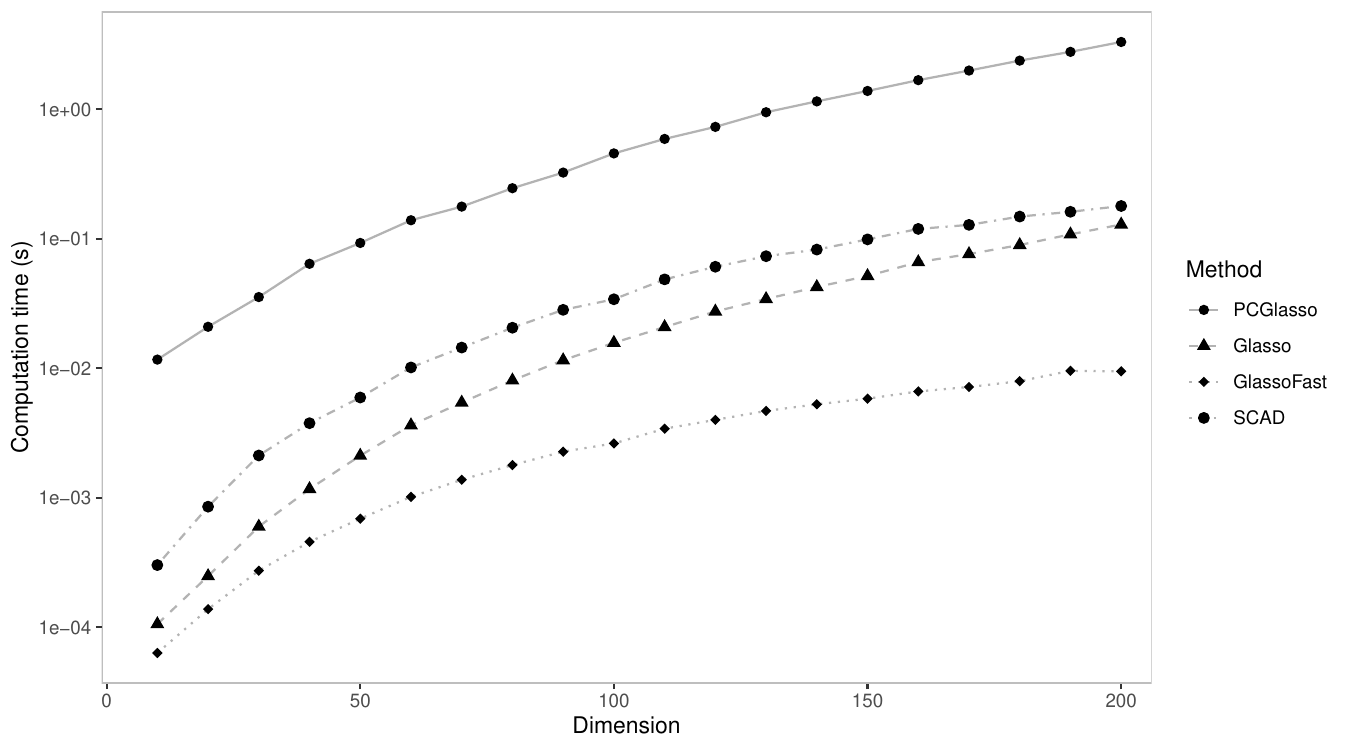} \\
	\includegraphics[scale=0.6]{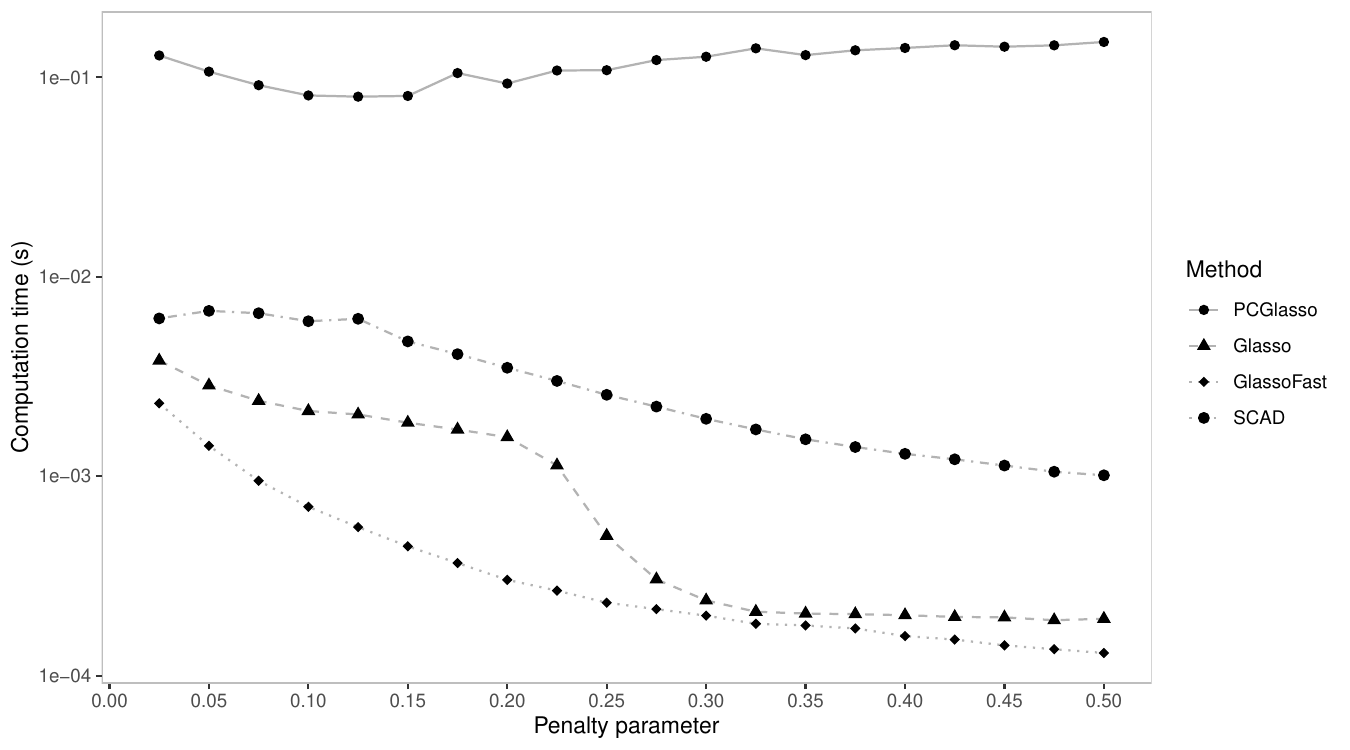} \\
	\includegraphics[scale=0.6]{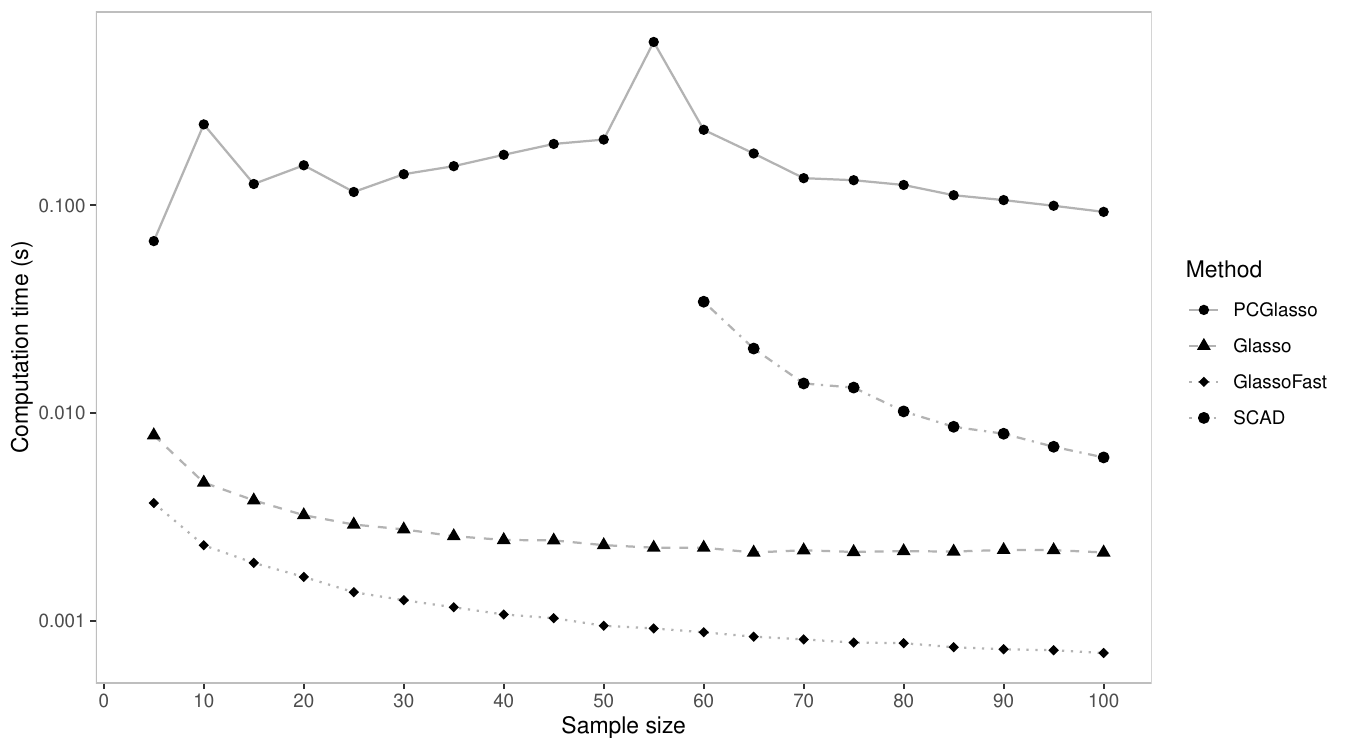}
\end{tabular}
\caption{Comparison of computation time for varying dimension (top), panalty parameter (middle) and sample size (bottom) for the hub graph.}
\label{fig:Hub}
\end{figure}

\begin{figure}[hp]
\centering
\begin{tabular}{c}
	\includegraphics[scale=0.6]{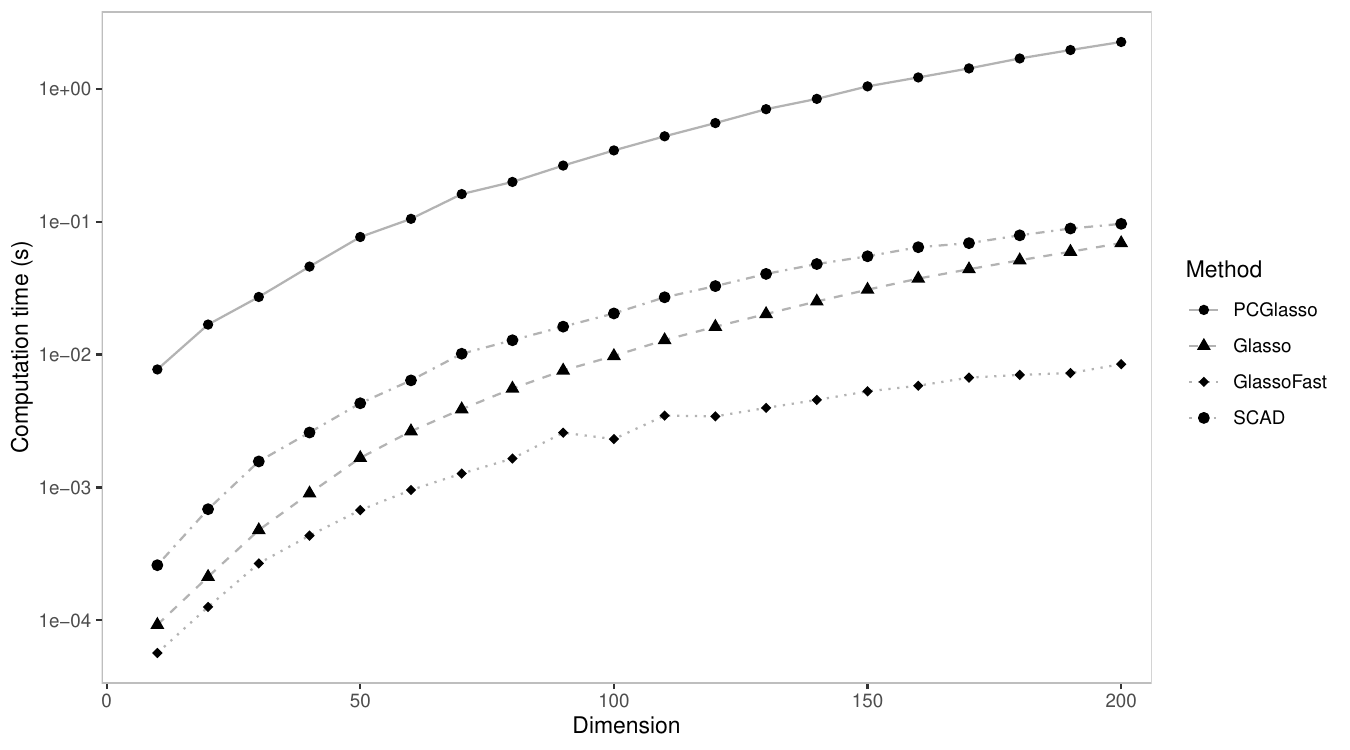} \\
	\includegraphics[scale=0.6]{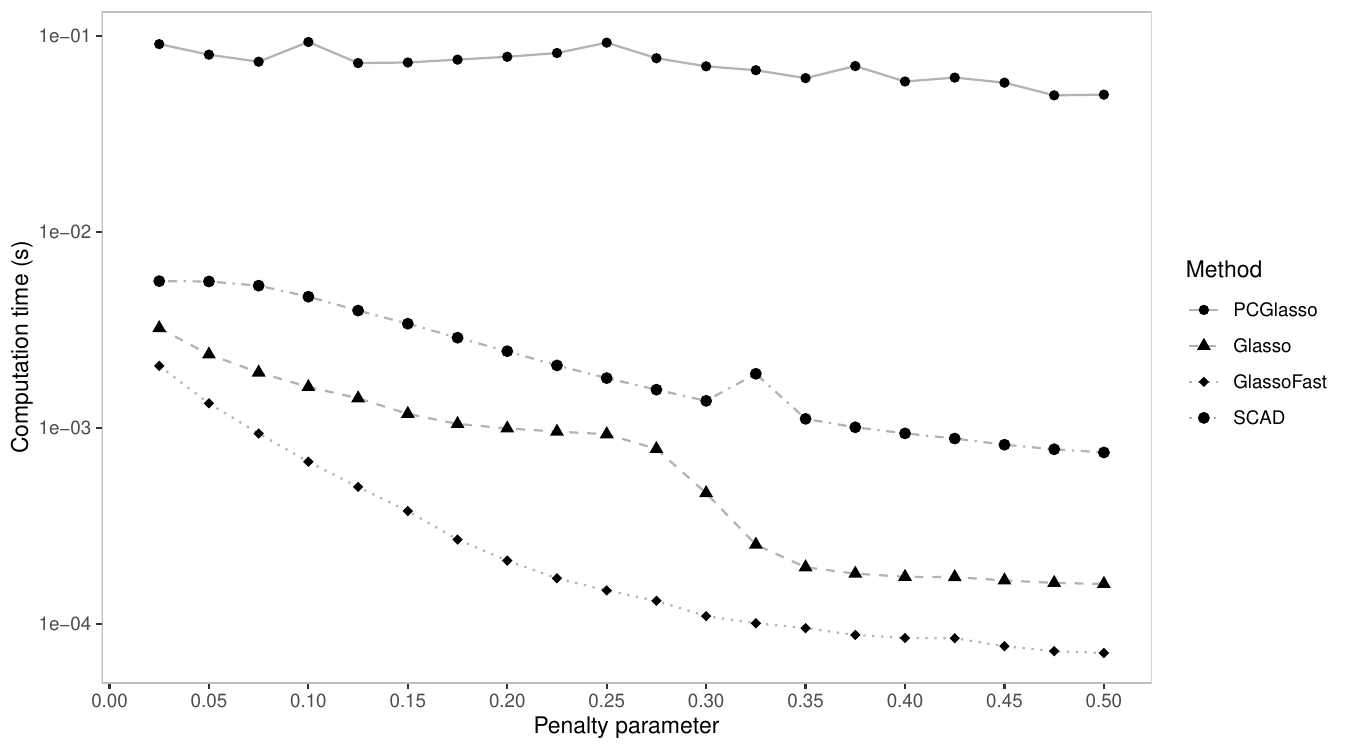} \\
	\includegraphics[scale=0.6]{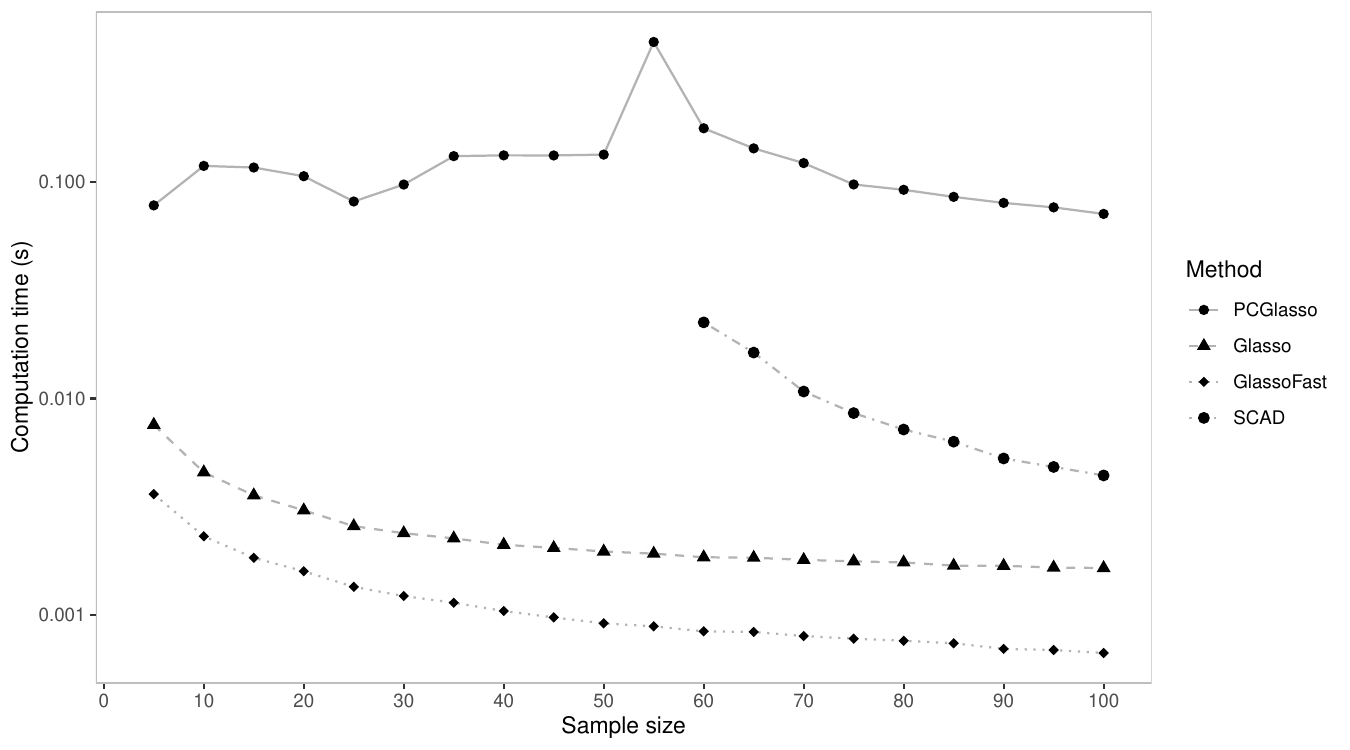}
\end{tabular}
\caption{Comparison of computation time for varying dimension (top), panalty parameter (middle) and sample size (bottom) for the AR2 model.}
\label{fig:AR2}
\end{figure}

\begin{figure}[hp]
\centering
\begin{tabular}{c}
	\includegraphics[scale=0.6]{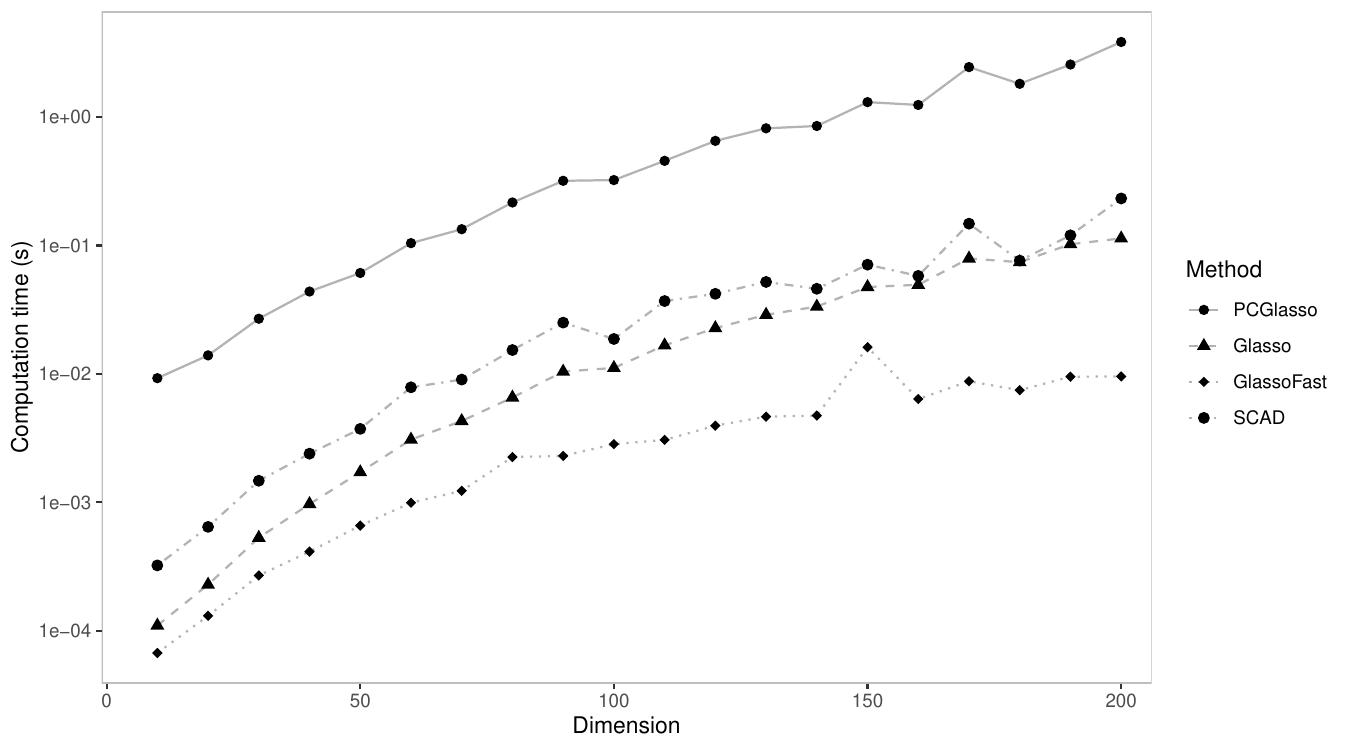} \\
	\includegraphics[scale=0.6]{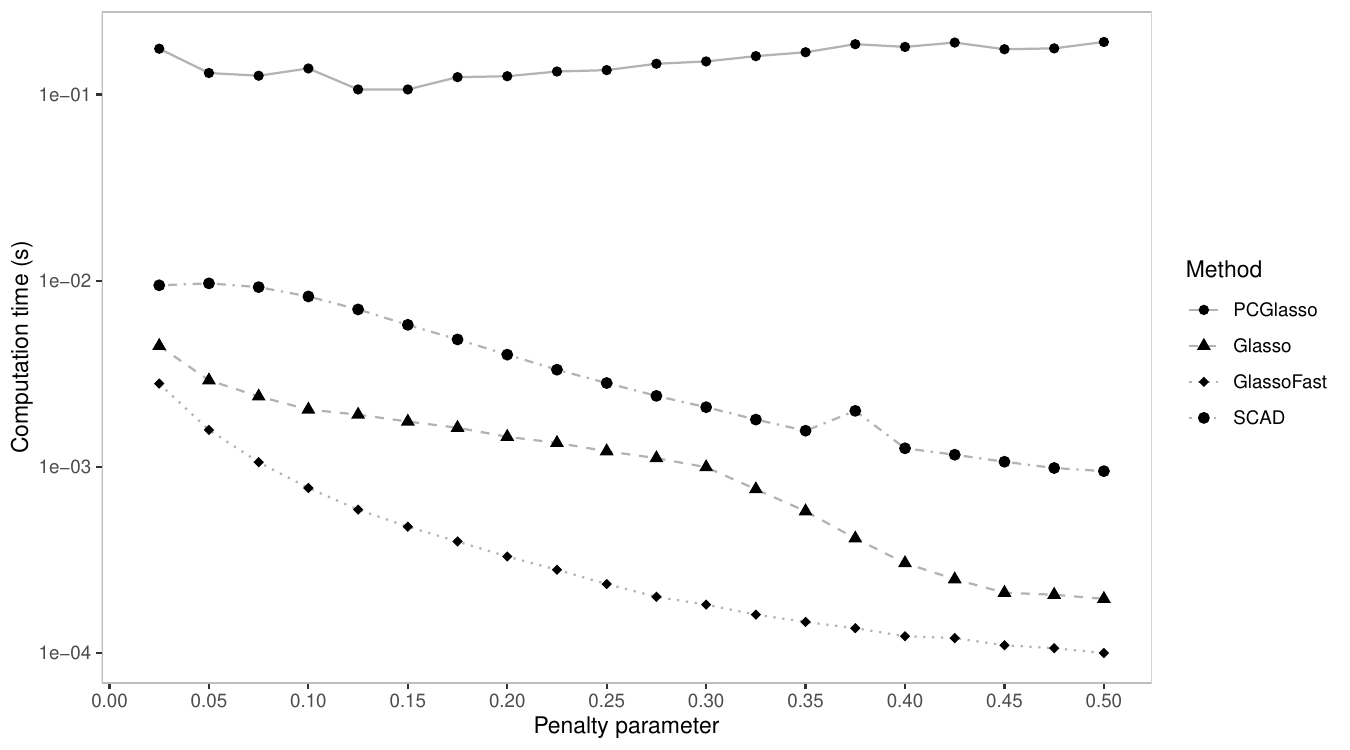} \\
	\includegraphics[scale=0.6]{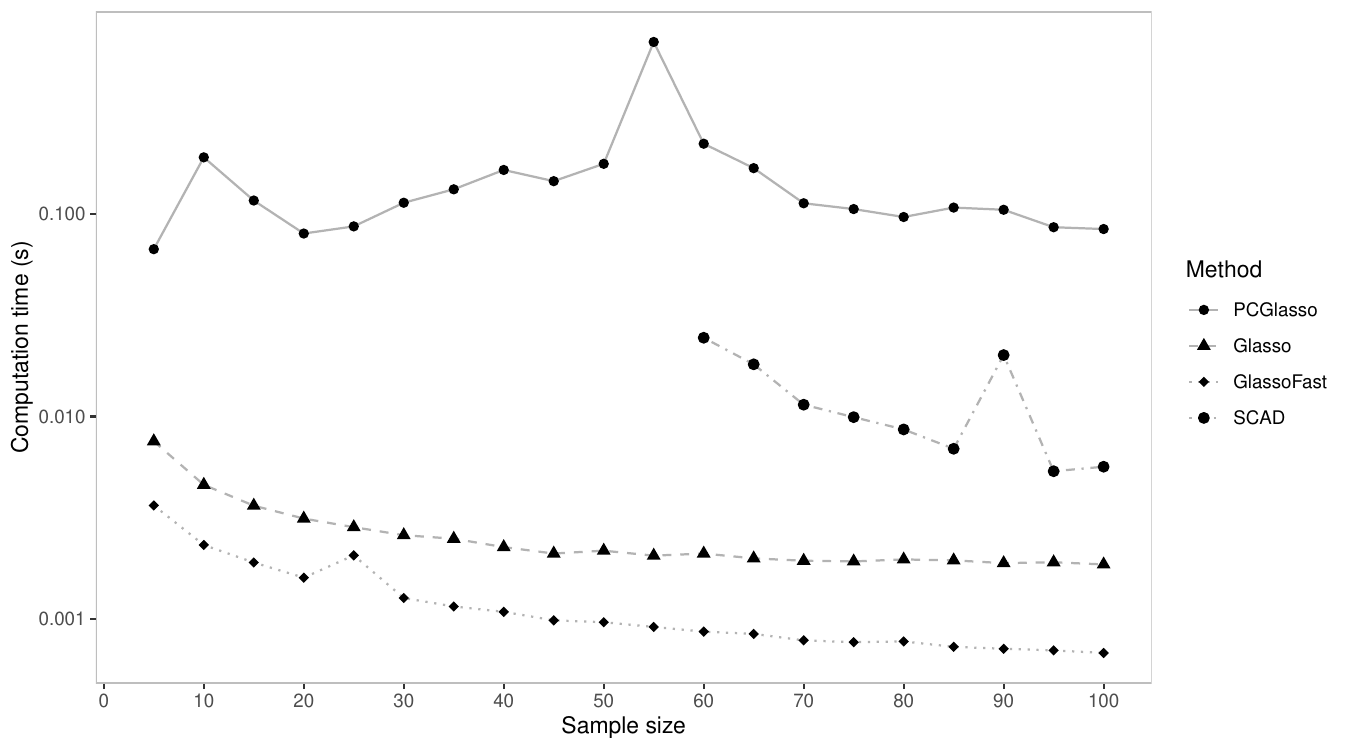}
\end{tabular}
\caption{Comparison of computation time for varying dimension (top), panalty parameter (middle) and sample size (bottom) for the random graph.}
\label{fig:Random}
\end{figure}

\end{document}